\documentclass[twocolumn,showpacs,preprintnumbers,amsmath,amssymb]{revtex4-1}

\usepackage{graphicx}
\usepackage{epstopdf} 

\usepackage{dcolumn}
\usepackage{bm}
\usepackage[active]{srcltx}
\usepackage{color}

\usepackage{multirow}
\usepackage{amsmath}
\usepackage{amssymb}
\usepackage{amsfonts}
\usepackage{amscd}
\usepackage{multirow}

\begin{document}

\title{Quantum dynamics of a dc-SQUID coupled to an asymmetric Cooper pair transistor}

\author{A. Fay$^1$,  W. Guichard$^1$, O. Buisson$^1$ and F. W. J. Hekking$^2$ }

\affiliation{$^1$Institut N\'eel, C.N.R.S.- Universit\'e
Joseph Fourier, BP 166, 38042 Grenoble-cedex 9, France}

\affiliation{$^2$LPMMC, C.N.R.S.- Universit\'e
Joseph Fourier, BP 166, 38042 Grenoble-cedex 9, France}


\begin{abstract}
We present a theoretical analysis of the quantum dynamics of a superconducting circuit based on a highly asymmetric Cooper pair
transistor (ACPT) in parallel to a dc-SQUID. Starting from the full Hamiltonian we show that the circuit can be modeled as  a charge qubit
(ACPT) coupled to an anharmonic oscillator (dc-SQUID). Depending on the anharmonicity of the SQUID, the Hamiltonian can be reduced either to one that describes
two coupled qubits or to the Jaynes-Cummings Hamiltonian. Here the dc-SQUID can be viewed as a tunable micron-size resonator. The coupling term,
which is a combination of a capacitive and a Josephson coupling between the two qubits, can be tuned from the very strong- to the zero-coupling regimes.
It describes very precisely the tunable coupling strength measured in this
circuit~\cite{Fay2008} and explains the `quantronium' as well as the adiabatic quantum transfer read-out.

\end{abstract}

\maketitle

\section{Introduction}

Two quantum systems with discrete energy levels coupled to each other form an elementary
block, useful for the study of fundamental phenomena and effects in quantum physics,
especially in the context of quantum information. The interaction between the two quantum
systems is essential to implement important concepts in this field such as entanglement,
quantum gate operations, and quantum information transfer. As to the theoretical
description of interacting quantum systems, two problems have been extensively studied in
particular: a two-level system (or qubit) coupled to a harmonic oscillator and two
coupled qubits. The former was used to describe among others the quantum electro-dynamics
associated with atoms in a cavity~\cite{Raimond_RMP01}, trapped ions coupled to their
vibrations~\cite{Leibfried_RMP03} and more generally interaction between matter and
photons~\cite{Tannoudji-Cohen_books}. The latter was developed to describe entangled
photons, trapped ions, and two-qubit quantum gate operations~\cite{Steane_RPP98}.

These studies were experimentally realized in the fields of quantum optics and atomic
physics and more recently extended to include quantum solid state devices. In particular,
during the last decade superconducting circuits have demonstrated their potential in
connection with quantum
experiments~\cite{Hofheinz2009,DiCarlo2009,Palacios-Laloy2010,Astafiev2010,Plantenberg2007,Wilson2007,Nakano2009,Palomaki2010,Hoskinson2009,Maklin_RMP01}.
They now appear as experimental model systems for studying fundamental quantum physics
and basic blocks for quantum information.

In this paper we consider a superconducting circuit composed of two well-known elements
coupled to each other: an inductive dc-SQUID and a Cooper pair transistor. This circuit
constitutes an elementary building block that can be operated in various parameter
regimes characterized by different types of quantum dynamics, as has been shown
experimentally in the past. As we will detail below, this is possible due to the fact
that three strongly coupled quantum variables determine the dynamics of this coupled
circuit.

For instance, when the current-biased dc-SQUID is non-inductive and classical and the
transistor is symmetric, one recovers the Quantronium~\cite{Vion_Science02}. When the
SQUID is operated in the nonlinear regime, the resulting system consists of a charge
qubit coupled to an anharmonic oscillator; this system has been shown to allow for
non-destructive quantum measurements~\cite{Boulant2007}. We recently operated the SQUID in
the quantum few-level limit~\cite{Fay2008}, demonstrating a very strong tunable coupling
between two different types of qubit: a phase qubit and a charge qubit.

The experimental performance of this circuit was limited by uncontrolled decoherence
sources. However its integration (three quantum variables strongly coupled on a
micrometer scale), its tunability, its various optimal points make this circuit
attractive once decoherence sources will be overcome upon technological improvements.

In this paper we present a rigorous theoretical analysis of the circuit in the parameter
range of our experiments~\cite{Fay2008}. The full Hamiltonian of the coupled circuit is
presented, describing a two-level system (Cooper pair transistor) coupled to two
anharmonic oscillators (dc-SQUID). In the relevant parameter range, the dc-SQUID dynamics
can be reduced from two-dimensional to one-dimensional. Consequently the dynamics of the
circuit is that of a qubit coupled to a single anharmonic oscillator. Depending on the
anharmonicity, different regimes can be studied in this unique circuit. When the
anharmonicity is neglected, we recover the physics of a qubit coupled to a harmonic
oscillator. Its quantum dynamics can then be described by the well-known Jaynes-Cummings
Hamiltonian. Although this limit can also be achieved with a qubit coupled to a high-Q
coplanar wave guide cavity~\cite{Wallraff2004,Hofheinz2009} we wish to emphasize that the use of a dc-SQUID
is of interest as it constitutes a micron-size resonator and it is tunable. When the oscillator
is considered anharmonic, its interaction with a qubit gives rise to very complex
dynamics which has been very little explored. If only the two lowest levels of the
oscillator are considered, it can be reduced to a two-level system. The coupled circuit
then describes two interacting qubits. In addition to the possibility to study different
dynamical regimes depending on the anharmonicity of the resonator, the coupling between
the SQUID and the transistor is fully tunable. As a result the system can be operated at
zero coupling, as well as in the weak and in the strong coupling limits.

In section II, after a description of the circuit under consideration, we construct the
Lagrangian and the Hamiltonian using Devoret's circuit theory~\cite{Devoret1995}.
Section III is devoted to the two-dimensional dynamics characteristic for an inductive
SQUID and its reduction to one-dimensional dynamics in the relevant parameter range. In
section IV we discuss the quantum dynamics of the asymmetric Cooper pair transistor,
especially at its two optimal points where it is insensitive to noise fluctuations. In
section V the terms describing the coupling between the dc-SQUID and the transistor are
discussed. In section VI, the full quantum dynamics of the coupled circuit is presented.
There we also discuss the two possible quantum measurements of the charge qubit that can
be performed by the dc-SQUID. The last section discusses the tunable coupling strength of
the circuit and its comparison with the experiments.

\section{Coupled circuit dynamics}

\subsection{Circuit description}

\begin{figure}
\includegraphics[width=0.45 \textwidth]{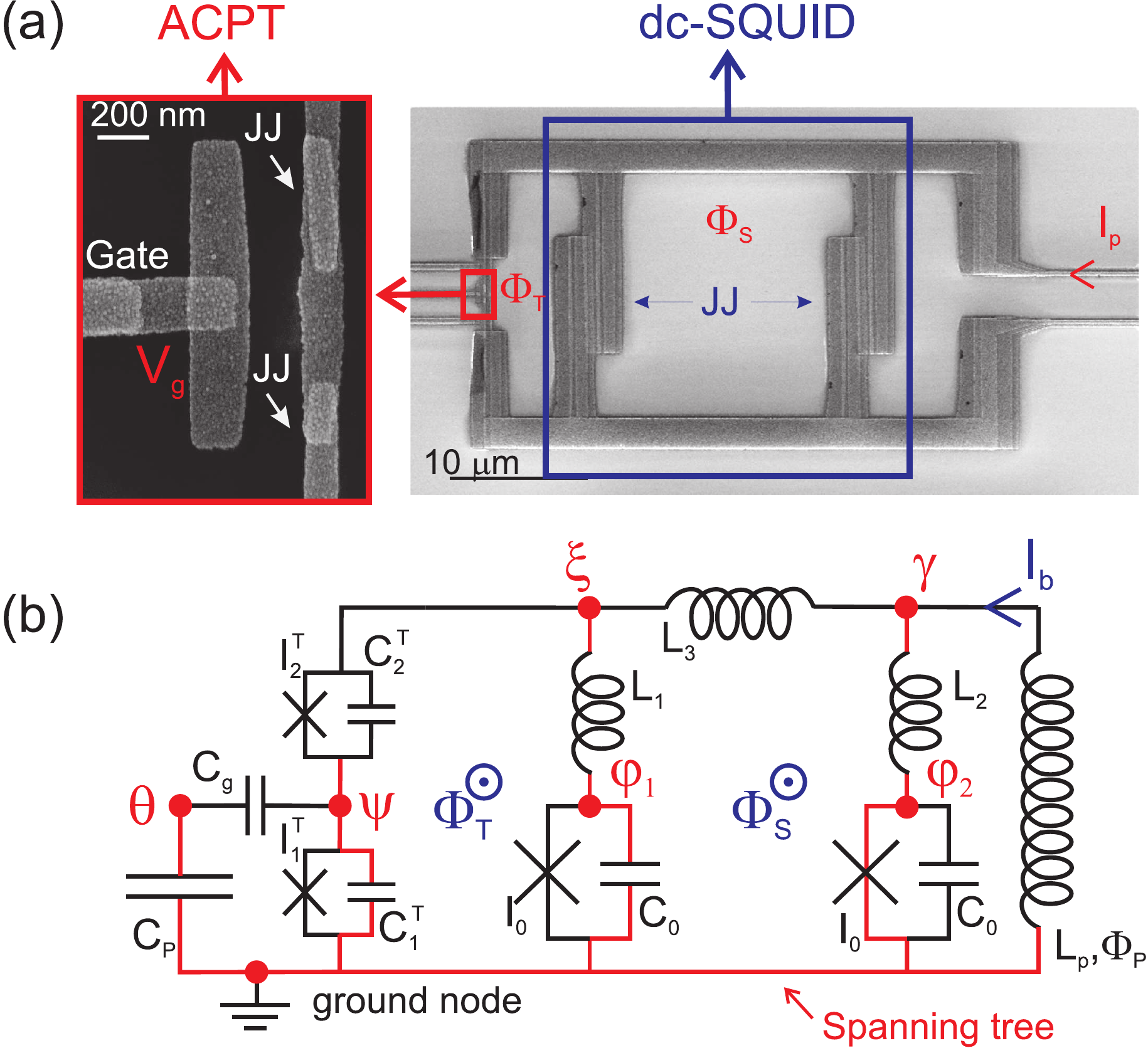}
\caption{The coupled circuit studied experimentally in Ref.~\cite{Fay2008}. (a) On the
right, a SEM picture shows the two elements of the circuit, i.e., an Asymmetric Cooper
Pair Transistor (red frame) connected in parallel to a dc-SQUID (blue frame). A zoom of
the transistor is shown on the left of the SEM picture. The transistor asymmetry stems
from the difference between the areas of the two Josephson junctions. The quantum
dynamics of the circuit can be modified by the bias current $I_b$, the flux $\Phi_S$ in
the dc-SQUID, the flux $\Phi_T$ in the other loop and the gate-voltage $V_g$. (b)
Schematic representation of the circuit. The four Josephson junctions are represented by
a pure Josephson element in parallel with a capacitance. The voltage bias of the
transistor gate and the current bias of the circuit are schematically indicated by an
infinite capacitance $C_P$ and an infinite inductance $L_P$ respectively. The spanning
tree drawn in red connects the ground node to the six active nodes $\varphi_1$,
$\varphi_2$, $\psi$, $\theta$, $\gamma$ and $\xi$.} \label{fig:schematic_circuit}
\end{figure}

A schematic electronic representation of the circuit studied theoretically hereafter is
presented in Fig.~\ref{fig:schematic_circuit}(b). In this circuit two different elements
are coupled that correspond to two basic blocks for typical superconducting quantum
devices. The first element is a dc-SQUID; it corresponds to a loop which contains two
identical Josephson junctions (JJ), each with a critical current $I_0$ and a capacitance
$C_0$. The total inductance $L_S$ of dc-SQUID is the sum of three inductances $L_1$,
$L_2$ and $L_3$ associated with the different parts of the SQUID loop. The second element
of the circuit is an asymmetric Cooper Pair transistor (ACPT) which is placed in parallel
with the dc-SQUID. The ACPT consists of a superconducting island connected to the
dc-SQUID by two different Josephson junctions. We denote by $I_1^T$ and $I_2^T$ the
critical currents and $C_1^T$ and $C_2^T$ the capacitances of these junctions. The
asymmetry of the transistor is characterized by the Josephson asymmetry parameter
$\mu=(I_2^T-I_1^T)/(I_2^T+I_1^T)$ and the capacitance asymmetry parameter
$\lambda=(C_2^T-C_1^T)/(C_2^T+C_1^T)$. The ACPT is also coupled to a gate-voltage; this
is modeled theoretically by an infinite capacitance $C_P$ with a charge $Q_b$ with $C_P,
Q_b \to \infty$ so the ratio $Q_b/C_b\equiv V_g$. The circuit is current-biased, modeled
similarly with the help of an infinite inductance $L_b \to \infty$ threaded by a flux
$\Phi_b \to \infty$ so that the ratio $\Phi_b/L_b\equiv I_b$ remains constant. The
properties of the circuit depend on four, experimentally tunable parameters $V_g$, $I_b$
and the fluxes $\Phi_S$ and $\Phi_T$ threading the dc-SQUID loop and the other loop of
the circuit, respectively. As we will see, these parameters allow to control and change
the dynamics of the coupled circuit. This circuit, realized and studied experimentally by
A. Fay \textit{et al.}~\cite{Fay2008}, is shown in the SEM view in
Fig.~\ref{fig:schematic_circuit}(a).

\subsection{Devoret's circuit theory}
The relevant degrees of freedom of a superconducting circuit and their dynamics can be
determined using the concept of node phases introduced by M. Devoret~\cite{Devoret1995}.
We distinguish two different kinds of nodes in the circuit. We first choose a ground node
to which the zero of phase is associated. Note that this choice corresponds to a choice
of gauge and is therefore arbitrary; although this choice affects the detailed form of
the Hamiltonian it does not affect the resulting dynamics of the circuit. The other nodes
of the circuit are called the active nodes, each described by an active phase. Six active
phases are present in the circuit considered here; they are denoted by $\varphi_1$,
$\varphi_2$, $\psi$, $\theta$, $\gamma$ and $\xi$.

Let us now introduce the so-called spanning tree~\cite{Devoret1995}. Starting from the
ground node we draw the branches of the spanning tree reaching each active node via a
unique path. In the case of the coupled circuit, the spanning tree (drawn in red in
Fig.~\ref{fig:schematic_circuit}(b)), connects the ground node (bottom node) to the six
active nodes.

\begin{figure}
\includegraphics[width=0.45 \textwidth]{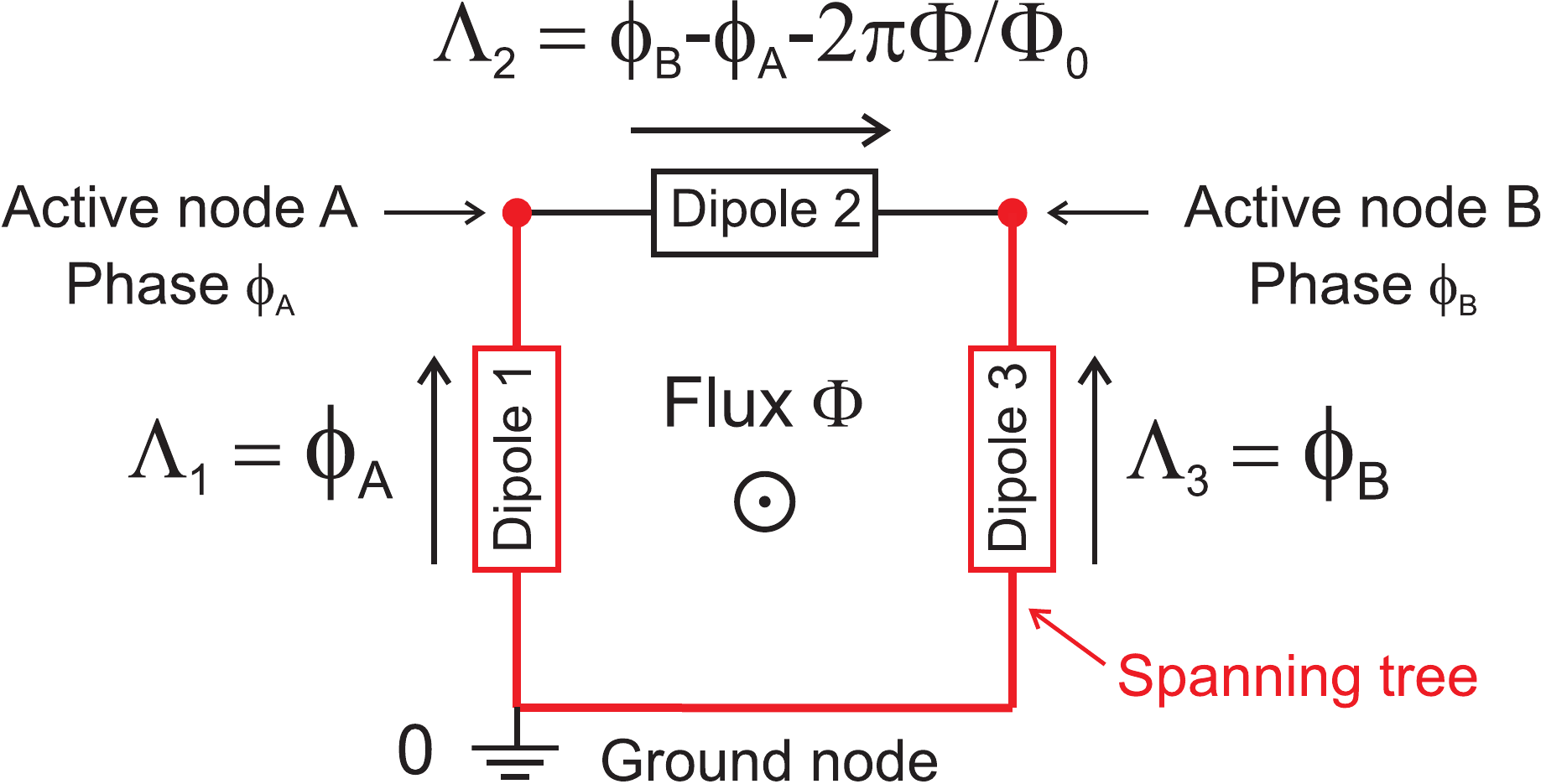}
\caption{Three dipoles in a loop threaded by the flux $\Phi$. The spanning tree in red
connects the ground node to the two active nodes $\phi_A$ and $\phi_B$. The
superconducting phases $\Lambda_1$, $\Lambda_2$ and $\Lambda_3$ across the three dipoles
respectively are functions of $\phi_A$, $\phi_B$ and $\Phi$.} \label{fig:simple_circuit}
\end{figure}

The superconducting phase difference across a dipole in a superconducting circuit can be
written as a function of the node phases with the help of two rules. We will illustrate
these rules with the help of the circuit presented in Fig.~\ref{fig:simple_circuit} as an
example. Here, three dipoles are placed in a loop threaded by the flux $\Phi$. In this
case, there are three nodes; the spanning tree (drawn in red) connects the ground node to
the two active nodes with phases $\phi_A$ and $\phi_B$ respectively. As a first rule, the
superconducting phase difference across a dipole located on the spanning tree is given by
the difference of the phases of the nodes linked to the dipole. Hence, the
superconducting phases $\Lambda_1$ and $\Lambda_3$ of the dipoles~1 and 3, respectively,
are given by $\Lambda_1=\phi_A$ and $\Lambda_2=\phi_B$. As a second rule, in the case of
a dipole which is not located on the spanning tree, we first define a minimal loop which
contains the previous dipole and the other dipoles located on the spanning tree. The
superconducting phase difference across the dipole is calculated, using the quantization
of the phase for the minimal loop \cite{Tinkham}. Let us apply this rule to determine the
superconducting phase difference $\Lambda_2$ across dipole~2. The minimal loop
corresponds simply to the loop of the circuit. The phase quantization in this loop gives
$\Lambda_2=\phi_B-\phi_A-2\pi\Phi/\Phi_0$, with $\Phi_0$ the superconducting flux
quantum. With the help of the two previous rules, we have analyzed the coupled circuit
and expressed the superconducting phase difference across each dipole as a function of
the six active phases.

\subsection{Current conservation and system dynamics}
\label{subsection:current_conservation} In a superconducting circuit, we find generally
three different dipoles: a capacitance $C$, an inductance $L$ and a Josephson element
$J$. The current through each of these dipoles can be expressed as a function of the
superconducting phase difference $\Lambda$ across the dipole. The voltage $V$ is related
to the phase difference $\Lambda$ by $V=\phi_0\dot\Lambda$. The current $I_{C}$ through
the capacitance is then given by $I_{C}=\phi_0 C \ddot\Lambda$, where
$\phi_0=\Phi_0/(2\pi)$. The current $I_L$ through the inductance reads
$I_L=\phi_0\Lambda/L$. Finally, the current through the Josephson element is given by the
Josephson relation $I_J=I_0\sin(\Lambda)$, where $I_0$ is the Josephson critical current
\cite{Josephson}. Using these expressions for the current, we can write the six current
conservation laws for each active node in the studied circuit (the sum of the currents
flowing into each node is zero). The six conservation laws yield six equations for the
dynamics of the node phases (cf. appendix \ref{appendix:current_conservation}), whose
solution yields the dynamics of the entire circuit.

\subsection{The Lagrangian}

The Lagrangian $\mathcal L$ of the circuit depends on the six node
phases and their time derivatives. Let us define the vector $\vec x
=(\varphi_1,\varphi_2,\psi,\theta,\gamma,\xi)$ formed by the six
node phases of the circuit. The six Euler-Lagrange equations are defined by \cite{Landau}:
\begin{equation}
\frac{{\rm d}}{{\rm d}t}\left(\frac{\partial \mathcal L}{\partial \dot{\vec
x}}\right)-\frac{\partial \mathcal L}{\partial \vec x}=0.
\end{equation}
The Lagrangian of the circuit has to be constructed in such a way that the Euler-Lagrange
equations are equivalent to the current conservation equations (Appendix
\ref{appendix:current_conservation}). We take the Lagrangian to be of the following form:
\begin{equation}
{\mathcal L(\dot{\vec x},\vec x)}=T(\dot{\vec x})-V(\vec x),
\label{eq:Lagrangien_T-V}
\end{equation}
with the kinetic energy
\begin{eqnarray}
{T(\dot{\vec x}})&=&\phi_0^2\big\{ \frac{1}{2} C_0\dot\varphi_1^2+
\frac{1}{2}
C_0\dot\varphi_2^2+\frac{1}{2}C_2^T(\dot{\xi}-\dot{\psi})^2\nonumber\\
&+&\frac{1}{2}C_1^T\dot{\psi}^2+\frac{1}{2}C_g(\dot\psi-\dot\theta)^2+\frac{1}{2}C_P\dot
\theta^2\big\}\label{eq:Lagrangien_T}
\end{eqnarray}
and the potential energy
\begin{eqnarray}
{V({\vec x}})&=&-\phi_0\big\{I_0\cos(\varphi_1)+I_0\cos(\varphi_2)\nonumber\\
&+&I_2^T\cos(\xi-\psi-\phi_T)+I_1^T\cos(\psi)\big\}\nonumber\\
&+&\phi_0^2\bigg\{\frac{1}{2}\frac{(\xi-\varphi_1)^2}{L_1}+\frac{1}{2}\frac{(\gamma-\varphi_2)^2}{L_2}\nonumber\\
&+&\frac{1}{2}\frac{(\gamma-\xi-\phi_S)^2}{L_3}-\frac{1}{\phi_0}I_p\,\gamma\bigg\}\label{eq:Lagrangien_V}.
\end{eqnarray}
The kinetic energy corresponds to the energy stored in the capacitances of the circuit,
whereas the potential energy is composed of the Josephson energies (cosine terms) and the
energy stored in the inductances of the circuit. From now on, we assume the SQUID
inductances $L_1=L_2$ and introduce the inductance asymmetry $\eta$ defined by $L_3=\eta
L_S$ \cite{remark}. Although the Lagrangian depends on six variables, the effective
low-energy behavior of the circuit is determined by three variables only, as we will see
below.

Let us first consider the phase variable $\gamma$. Its dynamics is that of a harmonic
oscillator of frequency  $\omega_\gamma = 1/\sqrt{\tilde{L}_S C_\gamma}$. Here $C_\gamma$
is the capacitance of the $\gamma$ node and $\tilde{L}_S = \eta (1-\eta) L_S /(1+\eta)$.
Using the circuit parameters of Ref.~\cite{Fay2008} (see Appendix
\ref{appendix:parameters}), where $C_\gamma$ is estimated to
be smaller than 0.1~fF, the frequency $\omega_\gamma$ is estimated to be larger than 1
THz, i.e., larger than all the other frequencies of the circuit. Next consider the phase
variable $\xi$. Again using the circuit parameters of Ref.~\cite{Fay2008}, we find that
the characteristic inductive currents $\phi_0/L_{1,2}$ are on the order of 1 $\mu A$,
much larger than the Josephson critical current  $I_2^T$ of about 1 nA. In other words,
the SQUID inductance is much smaller than the Josephson inductance $\sim 1/I_2^T$.
Therefore, in a first approximation, the dynamics of $\xi$ is also of the harmonic
oscillator type with a frequency $\omega_\xi = \sqrt{1/\tilde{L}_S C_2^T}$. We estimate
$\omega_\xi$ to be around 640 GHz. This frequency is much smaller than $\omega_\gamma$,
but still quite high compared to the frequencies characterizing the dynamics of the
variables $\varphi_1$, $\varphi_2$, $\psi$ and $\theta$ (around 10 GHz, see below). This
implies that we can use the adiabatic approximation to eliminate the fast variables
$\gamma$ and $\xi$ and obtain an effective Lagrangian for the slow variables $\phi_1,
\phi_2, \xi$, and $\theta$.

In order to implement the adiabatic approximation, we write $\gamma  = \gamma_0 + \delta
\gamma$ and similarly $\xi = \xi_0 + \delta \xi$. Here $\gamma_0$ and $\xi_0$ correspond
the minima of the harmonic potentials confining the motion of these variables,
\begin{eqnarray}
\gamma_0(x,y) &=& x+\eta
y+\frac{\phi_S}{2}(1-\eta)+\frac{1}{4\phi_0}L_S
I_p(1-\eta^2),\label{eq:gamma_function_xy}\\
\xi_0(x,y) &=& x-\eta y
-\frac{\phi_S}{2}(1-\eta)+\frac{1}{4\phi_0}L_SI_p(1-\eta)^2\label{eq:xi_function_xy},
\end{eqnarray}
where $x=(\varphi_1+\varphi_2)/2$ and $y=(\varphi_2-\varphi_1)/2$. Note that $\gamma_0$
and $\xi_0$ depend on $x$ and $y$, hence they acquire slow dynamics through the phases
$\varphi_1$ and $\varphi_2$. The dynamics of the deviations $\delta \gamma$ and $\delta
\xi$ is fast, determined by the frequencies $\omega_\gamma$ and $\omega_\xi$,
respectively. Substituting the above decomposition for $\gamma$ and $\xi$ in the
Lagrangian (\ref{eq:Lagrangien_T-V}), (\ref{eq:Lagrangien_T}), and
(\ref{eq:Lagrangien_V}), and averaging over the fast variables $\delta \gamma$ and
$\delta \xi$ we find the effective low-frequency potential energy
\begin{eqnarray}
V(x,y,\psi)=\nonumber \\2E_J \left(-\cos(x)\cos(y)-s(\eta y+
x)+b{(y-y_B)^2}\right)\nonumber\\
-{E_{J1}^T}\cos(\psi)-{E_{J2}^T}\cos\left(\psi-\xi_0(x,y)
+\phi_T\right).\label{eq:Lagrangien_V_xy}
\end{eqnarray}
Here we defined the reduced parameters $b=\phi_0/(LI_0)$, $s=I_b/(2I_0)$ and
$y_B=\pi/(\Phi_S/\Phi_0)$. Furtermore, $E_J=\phi_0 I_0$, ${E_{J1}^T}=\phi_0 I_1^T$ and
${E_{J2}^T}=\phi_0 I_2^T$ are the different Josephson energies of the circuit. Note that
the fast oscillations of $\xi$ lead to a renormalization of $I_2^T$; the value it takes
in the effective low-frequency potential (\ref{eq:Lagrangien_V_xy}) will therefore be
smaller than the bare value appearing in (\ref{eq:Lagrangien_V}). Assuming that the bias
current $I_b$ and the flux $\Phi_S$ are constant, we deduce from
Eqs.~(\ref{eq:gamma_function_xy}) and (\ref{eq:xi_function_xy}) that  $\dot\gamma_0=\dot
x+\eta \dot y$ and $\dot\xi_0=\dot x-\eta \dot y$. The kinetic part of the Lagrangian can
then be expressed as:
\begin{eqnarray}
T(\dot x,\dot y,\dot\psi,\dot\theta)&=&\phi_0^2\bigg\{C_0\dot
x^2+C_0\dot y^2+\frac{1}{2}C_2^T(\dot x-\eta\dot
y-\dot{\psi})^2\nonumber\\
&+&\frac{1}{2}C_1^T\dot{\psi}^2+\frac{1}{2}C_g(\dot\psi-\dot\theta)^2+\frac{1}{2}C_P\dot
\theta^2\bigg\}. \label{eq:Lagrangien_T_2}
\end{eqnarray}
The four variables of the Lagrangian can be separated in three groups. Indeed, the
dynamics of $x$ and $y$ corresponds to that of the dc-SQUID, whereas the dynamics of
$\psi$ is associated with that of the ACPT. The variable $\theta$ is used to model the
effect of the gate voltage (cf. Sec.~\ref{subsec:Hamiltonian}). Note that the last term
of the potential~(\ref{eq:Lagrangien_V_xy}) and the third term of the kinetic term~
(\ref{eq:Lagrangien_T_2}) couple the variables of the dc-SQUID and those of the ACPT
together, and therefore are responsible of the coupling between these two elements.

\subsection{Choice of variables for the dc-SQUID}
\label{subsec:variables_SQUID}

\begin{figure}
\centering
\includegraphics[width=0.8\linewidth]{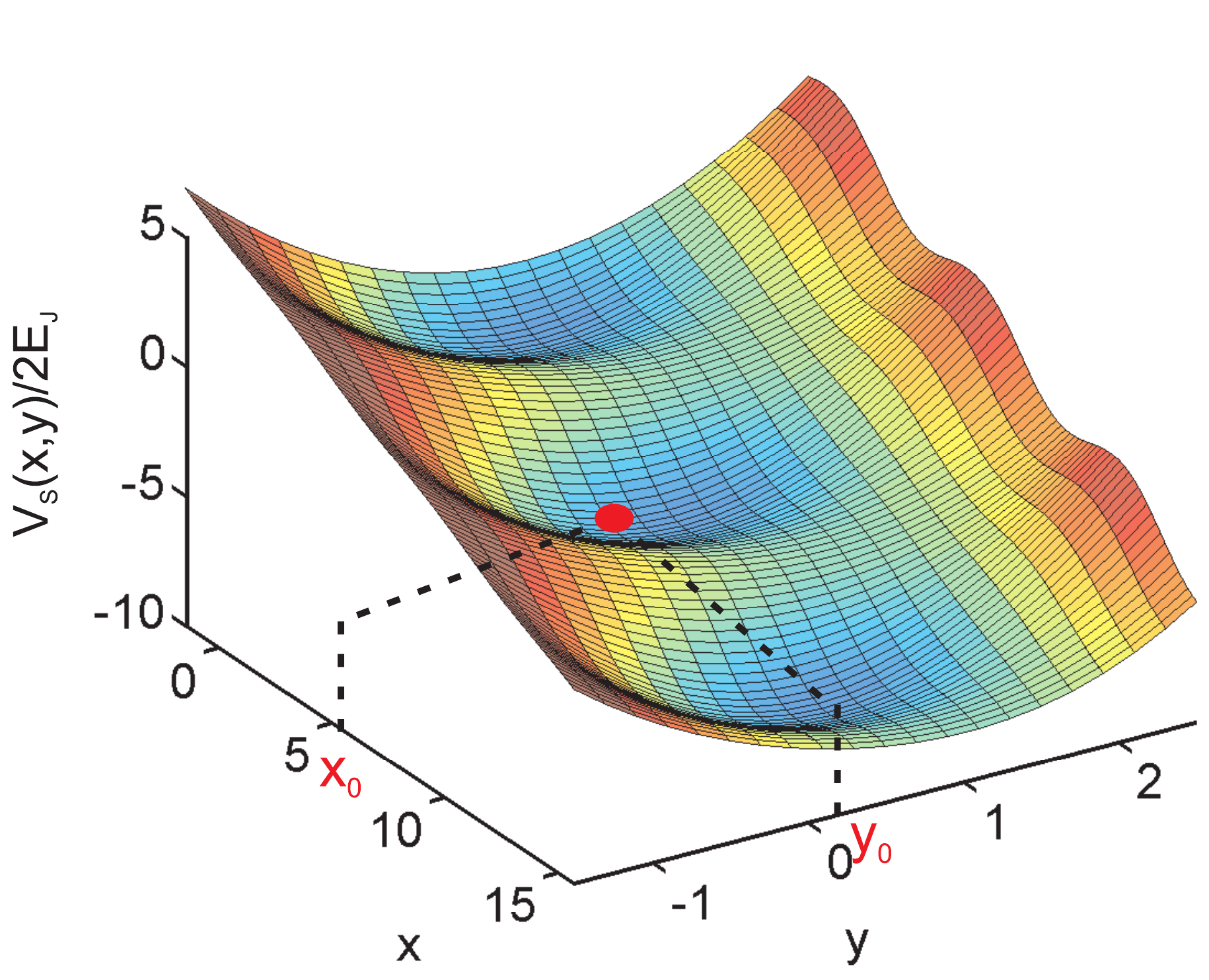}
\includegraphics[width=0.7\linewidth]{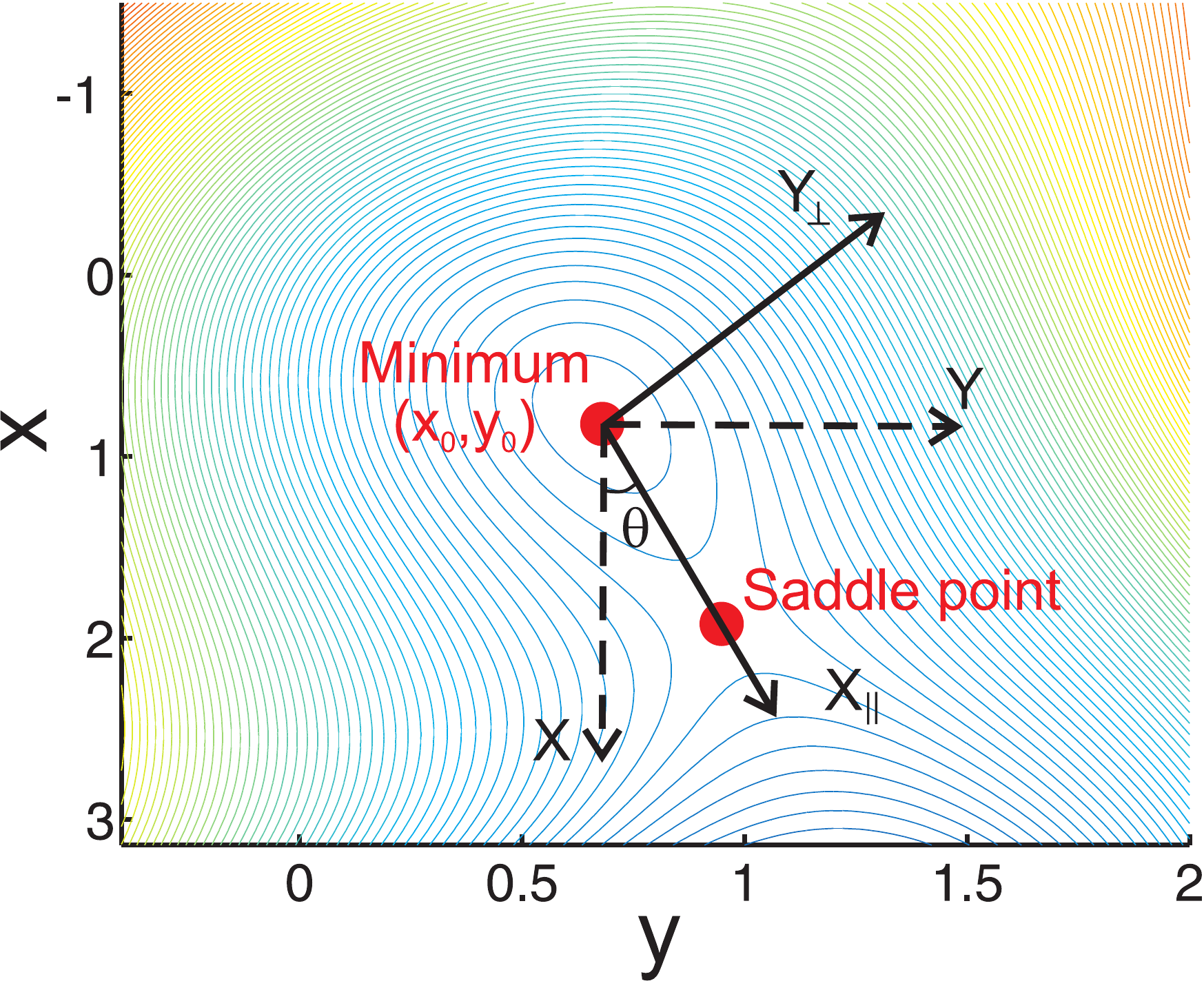}
\caption{Potential $V_{S}(x,y)$ of the dc-SQUID with bias parameters $s=0.55$,
$y_B=0$ (SQUID parameters $b=1.28$, $\eta=0.29$). The point ($x_0,y_0$) is the local minimum of one of the potential wells.} \label{fig:V_SQUID}
\end{figure}

The Lagrangian of the circuit is a function of the variables $x$ and $y$ associated with
the SQUID. We want to change these variables to more appropriate ones which will be used
below to describe the dynamics of the dc-SQUID (cf.~Sec.~\ref{sec:SQUID}). Let us
introduce the two-dimensional potential of the dc-SQUID $V_S(x,y)$ which is the
contribution to the potential $V(x,y,\psi)$, Eq.~(\ref{eq:Lagrangien_V_xy}), depending
only on the variables $x$ and $y$. It reads:
\begin{equation}
V_{S}(x,y)=2E_J \left(-\cos(x)\cos(y)-s(\eta y+
x)+b(y-y_B)^2\right).\label{eq:V_SQUID}
\end{equation}
We stress here that this potential is identical to the one of a dc-SQUID alone, as
studied by J. Claudon \emph{et al.} \cite{Claudon2004}. The dynamics of the dc-SQUID is
similar to that of a fictitious particle of mass $\approx \phi_0^2 2 C_0$ which evolves
in the potential $V_{S}(x,y)$. This potential undulates due to the product of cosine
terms and contains wells that are separated by saddle points (see Fig \ref{fig:V_SQUID});
$V_{S}(x,y)$ is modulated by the bias current $I_b$ and the flux $\Phi_S$. We consider
now the case that the particle is trapped in one of these wells associated with a given
local minimum $(x_0,y_0)$. Let us introduce the displacement variables around $(x_0,y_0)$
defined by $X=x-x_0$ and $Y=y-y_0$. We assume that the particle's motion does not extend
far from $(x_0,y_0)$. Then, we can replace the potential $V_{S}(x,y)$ by its third order
expansion around $(X=0,Y=0)$. This expansion contains a cross-term in $XY$ which
disappears by performing a rotation of the (X,Y) plane (Fig.~\ref{fig:V_SQUID}(b)) by the
angle $\theta$, where $\theta$ is given by
$
\tan(2\theta)/2={\partial_{xy}^2U(x_0,y_0)}/({\partial_{xx}^2U(x_0,y_0)-\partial_{yy}^2U(x_0,y_0)}).
$
The new variables of the dc-SQUID $X_\parallel$ and $Y_\perp$ associated with the rotated plane are
defined by
\begin{equation}
\left(\begin{array}{c}
X_\parallel\\
Y_\perp
\end{array}\right)=
\left(
\begin{array}{cc}
\cos(\theta) & \sin(\theta) \\
-\sin(\theta) & \cos(\theta)
\end{array}\right)
\left(\begin{array}{c}
X\\
Y
\end{array}\right)\label{eq:X_to_Xpar}
\end{equation}
and correspond to the position of the particle along the longitudinal and the transverse
direction, defined by the minimal and the maximal curvature of the potential,
respectively. The third-order expansion of the SQUID potential now takes the form
\begin{eqnarray}
V_S(X_\parallel,Y_\perp)&=&\left[\frac{1}{2}k_\parallel
X_\parallel^2+\tilde{\sigma}_\parallel X_\parallel^3\right]+
\left[\frac{1}{2}k_\perp Y_\perp^2+\tilde{\sigma}_\perp
Y_\perp^3\right]\nonumber\\&+&\left[\beta_a  Y_\perp^2 X_\parallel +
\beta_b Y_\perp X_\parallel^2\right],
\end{eqnarray}
where the prefactors $k_\parallel$, $\tilde{\sigma}_\parallel$, $k_\perp$,
$\tilde{\sigma}_\perp$, $\beta_a$ and $\beta_b$ can be calculated numerically. The
dynamics of the fictitious particle of the SQUID in this potential will be analyzed in
Sec.~\ref{sec:SQUID}.
%
The full potential appearing in the Lagrangian, Eq.~(\ref{eq:Lagrangien_V_xy}), is given
by:
\begin{eqnarray}
&&\!\!\!\!\!\!\!\!\!\!\!\!\!\!V(X_\parallel,Y_\perp,\psi)=\nonumber\\
&&\left[\frac{1}{2}k_\parallel
X_\parallel^2+\tilde{\sigma}_\parallel X_\parallel^3\right]+
\left[\frac{1}{2}k_\perp Y_\perp^2+\tilde{\sigma}_\perp
Y_\perp^3\right]\nonumber\\&&+\left[\beta_a Y_\perp^2 X_\parallel +
\beta_b
Y_\perp X_\parallel^2\right]  \nonumber\\
&&- {E_{J1}^T} \cos(\psi) -{E_{J2}^T} \cos(\hat \psi - \delta)\nonumber\\
&&- {E_{J2}^T} \alpha_\parallel \hat X_\parallel
\sin(\hat\psi-\delta)+{E_{J2}^T}\alpha_\perp \hat Y_\perp
\sin(\hat\psi-\delta),\label{eq:pot_lagrangien}
\end{eqnarray}
where $\delta\equiv\xi(x_0,y_0)-\phi_T$ is the classical phase difference across the
transistor. The prefactors $\alpha_\parallel\equiv\cos(\theta)-\eta\sin(\theta)$ and
$\alpha_\perp\equiv\sin(\theta)+\eta\cos(\theta)$ reflect the two-dimensionality of the
SQUID potential. When the inductance $L_S$ is zero, the dynamics of the SQUID is
described by only one variable, $x$, since $y=y_B$. In that case, the prefactors
$\alpha_\parallel$ and $\alpha_\perp$ are equal to one. The two last terms of the
potential contain one variable of the SQUID and one of the transistor. They thus couple
the SQUID and the transistor. Note that the coupling terms of the second order and beyond
have been neglected in the potential. The kinetic term of the Lagrangian can be rewritten
as, using the variables $X_\parallel$ and $Y_\perp$,
\begin{eqnarray}
&&T(\dot X_\parallel,\dot Y_\perp,\dot \psi,\dot
\theta)=\nonumber\\
&\ &\frac{1}{2}\phi_0^2\bigg\{(2C_0+\alpha_\parallel^2C_2^T)\dot{X}_\parallel^2
+(2C_0+\alpha_\perp^2C_2^T)\dot{Y}_\perp^2+C_\Sigma\dot{\psi}^2\nonumber\\
&\  &-2C_2^T\left(\alpha_\parallel\dot{X}_\parallel-\alpha_\perp\dot{Y}_\perp
\right)\dot\psi
-2C_2^T\alpha_\parallel\alpha_\perp\dot{X}_\parallel\dot{Y}_\perp\nonumber\\
&\  &-2C_g\dot{\psi}\dot\theta+(C_P+C_g)\dot{\theta}^2 \bigg\},
\label{eq:lagrangien_cinetique}
\end{eqnarray}
where $C_\Sigma$ is the total capacitor of the transistor, defined by
$C_\Sigma=C_1^T+C_2^T+C_g$. The final expression for the total Lagrangian is obtained
from $\mathcal{L}(\dot X_\parallel,\dot Y_\perp,\dot \psi,\dot
\theta,X_\parallel,Y_\perp,\psi)=T(\dot X_\parallel,\dot Y_\perp,\dot \psi,\dot
\theta)-V(X_\parallel,Y_\perp,\psi)$; it will be used in the next section to establish
the Hamiltonian of the circuit.

\subsection{The Hamiltonian}
\label{subsec:Hamiltonian} The Hamiltonian of the coupled circuit is a function of the
variables $(X_\parallel,Y_\perp,\psi,\theta)$ and the conjugate momenta $(-\hbar
P_\parallel,-\hbar P_\perp,-\hbar n,-\hbar n_Q)$. These momenta are related to the
velocities $(\dot X_\parallel,\dot Y_\perp,\dot \psi,\dot\theta)$ by the well-known
expressions $-\hbar P_\parallel\equiv\partial {\mathcal L}/\partial \dot X_\parallel$,
$-\hbar P_\perp\equiv\partial {\mathcal L}/\partial\dot Y_\perp$, $-\hbar n\equiv\partial
{\mathcal L}/\partial \dot\Psi$ and $-\hbar n_Q\equiv\partial {\mathcal L}/\partial
\dot\theta$. The analytical expressions for the conjugate momentum variables are given in
appendix \ref{appendix:conjugate_variables}. The conjugate variables generate the charges
$P_\parallel$, $n$ and $n_Q$ with unit [-2e]. We stress that these charges have a clear
physical meaning. Indeed, $P_\parallel$ corresponds to the number of Cooper pairs stored
in the two capacitors $C_0$; $n$ is the number of Cooper pairs on the island. In
Eq.~(\ref{eq:nq}), the charge $-2e n_Q$ is equal to the bias charge $Q_b$. Performing the
limiting procedure $C_P,Q_b \to \infty$, keeping their ratio constant $Q_b/C_P = V_g$, we
see that the velocity $\dot\theta$ is constant and defined by $\dot\theta=V_g/\phi_0$.
The expression for the Hamiltonian is determined by the Legendre transformation
\cite{Landau}:
\begin{equation}
{\mathcal H}= -\hbar P_\parallel \dot{X}_\parallel-\hbar P_\perp
\dot{Y}_\perp - \hbar n \dot{\psi}-{\mathcal L}.\label{eq:base_hamiltonian}
\end{equation}
All the velocities which appear in the Lagrangian have to be replaced by the conjugate
variables, inverting the set of equations given in appendix
\ref{appendix:conjugate_variables}.
The full Hamiltonian of the circuit then takes the form:
\begin{eqnarray}
&& \!\!\!\!\!\!\!\!\!{\widehat
H}({\color{blue}\widetilde{{P}}_\parallel},{\color{blue}\widehat{X}_\parallel},{\color{green}\widetilde{{P}}_\perp},{\color{green}\widehat{Y}_\perp},{\color{red}\hat{n}},{\color{red}\hat\psi})=\nonumber\\
&&\frac{(2e)^2}{2C_\parallel}{\color{blue}\widetilde{{P}}_\parallel^2}
+ \frac{1}{2}k_\parallel
{\color{blue}\widetilde{X}_\parallel^2}+\tilde{\sigma}{\color{blue}\widetilde{{X}}_\parallel^3}\nonumber\\
&\ &+\frac{(2e)^2}{2C_\perp}{\color{green}\widetilde{{P}}_\perp^2}+\frac{1}{2}k_\perp
{\color{green}\widetilde{{Y}}_\perp^2}+\tilde{\sigma}_\perp
{\color{green}\widetilde{{Y}}_\perp^3}\nonumber\\
&\ &+\frac{(2e)^2}{C_{\parallel\perp}}{\color{blue}\widetilde{{P}}_\parallel}{\color{green}\widetilde{{P}}_\perp}+ \beta_a  {\color{green}\widetilde{{Y}}_\perp^2}{\color{blue}\widetilde{{X}}_\parallel} + \beta_b {\color{green} \widetilde{{Y}}_\perp}{\color{blue}\widetilde{{X}}_\parallel^2}\nonumber\\
&\ &+\frac{(2e)^2}{2C_T}\left({\color{red}\hat{n}}-\frac{C_gV_g}{2e}\right)^2- {E_{J1}^T} \cos({\color{red}\hat\psi}) -{E_{J2}^T} \cos({\color{red}\hat \psi} - \delta)\nonumber\\
&\ &+\frac{(2e)^2}{C_{n\parallel}}{\color{blue}\widetilde{{P}}_\parallel}\left({\color{red}\hat{n}}-\frac{C_gV_g}{2e}\right)-
{E_{J2}^T} \alpha_\parallel {\color{blue}\widetilde{X}_\parallel}
\sin({\color{red}\hat\psi}-\delta)\nonumber\\
&\ &-\frac{(2e)^2}{C_{n\perp}}{\color{green}\widetilde{{P}}_\perp}\left({\color{red}\hat{n}}-\frac{C_gV_g}{2e}\right)+{E_{J2}^T}\alpha_\perp
{\color{green}\widetilde{Y}_\perp}
\sin({\color{red}\hat\psi}-\delta), \label{eq:Hamiltonien_total}
\end{eqnarray}
where the analytical expressions of the capacitances $C_\parallel$, $C_\perp$, $C_T$,
$C_{n\parallel}$, $C_{n\perp}$ and $C_{\parallel\perp}$ are given in
Tab.~\ref{tab:capacitances}. Applying the standard canonical quantization rules, the
classical variables have been replaced by their corresponding quantum operators. For more
clarity, the conjugate pairs appear in the Hamiltonian in different colors. They satisfy
the following commutation operations
\begin{equation}
\left\{
\begin{aligned}
& [{\color{blue}\widetilde{X}_\parallel},{\color{blue} \widetilde{P}_\parallel}]=-i \\
& [{\color{green}\widetilde{Y}_\perp},{\color{green} \widetilde{
P}_\perp}]=-i\\
& [{\color{red}\hat\psi},{\color{red}\hat n}]=-i
\end{aligned}\right.
\end{equation}
\begin{table*}[htb!]
\centering
\begin{tabular}{c c c c c c c}
\hline\hline
capacitance labels & $C_\parallel$ & $C_\perp$ & $C_T$ & $C_{n\parallel}$
& $C_{n\perp}$ & $C_{\parallel\perp}$\\
\hline
exact expressions & $\frac{\bar{C}^2}{C_\Sigma+\alpha_\perp C_R}$ &
$\frac{\bar{C}^2}{C_\Sigma+\alpha_\parallel C_R}$ & $\frac{\bar{C}^2}{2C_0 + C_2^T
(1+\eta^2)}$ & $\frac{\bar{C}^2}{\alpha_\parallel C_2^T}$
& $\frac{\bar{C}^2}{(\alpha_\perp C_2^T)}$ & $\frac{\bar{C}^2}{(\alpha_\perp \alpha_\parallel C_R)}$\\
approximated expressions & $2C_0$ & $2C_0$ & $C_1^T+C_2^T$ &
$\frac{2C_0(C_2^T+C_1^T)}{\alpha_\parallel C_2^T}$
& $\frac{2C_0(C_2^T+C_1^T)}{\alpha_\perp C_2^T}$ & $\frac{\bar{C}^2}{\alpha_\perp \alpha_\parallel C_R}$\\
numerical values & $455.6$ fF & $455.1$ fF & $2.90$ fF & $653.0$ fF
& $2.252$ pF & $1.17$ nF\\
\hline\hline
\end{tabular} \caption{Analytical expressions for the capacitances present in Hamiltonian (\ref{eq:Hamiltonien_total})
and the approximated expressions in the limit of $C_g\ll C_2^T\ \&\ C_1^T\ll C_0$. We have
used the definitions $\bar{C}^2=2C_0(C_\Sigma+C_R(1+\eta^2))$, $C_\Sigma=C_1^T+C_2^T+C_g$
and $C_R=C_2^T(C_1^T+C_g)/2C_0$. The numerical values have been
calculated for the parameters of the circuit studied
in Ref.~\cite{Fay2008}: $C_0=227$ fF, $C_1^T=2.0$ fF, $C_2^T=0.9$ fF, $C_g=29$ aF,
and for a zero escape angle ($\theta=0$).} \label{tab:capacitances}
\end{table*}

The properties of the Hamiltonian (\ref{eq:Hamiltonien_total}) are not trivial. It
describes the quantum dynamics of three sub-systems: the longitudinal and transverse
phase oscillations within the dc-SQUID and the charge dynamics of the ACPT. Moreover the
Hamiltonian describes the dominant coupling between the different quantum sub-systems.
Very complex dynamics can appear in this full circuit. In this paper we
mainly concentrate on the dynamics of the longitudinal SQUID phase mode and the charge dynamics
of the ACPT and as well as on their coupling.
The next section is dedicated to the study of the dc-SQUID Hamiltonian.
We will deduce the simplified Hamiltonian for the longitudinal phase mode and
justify why transverse phase mode can be neglected in this study.

\section{dc-SQUID}
\label{sec:SQUID} The dc-SQUID potential has already been discussed in
Sec.~\ref{subsec:variables_SQUID}, where we introduced the change of the variables $x$
and $y$ to the variables $X_\parallel$ and $Y_\perp$. In this section, we first analyze
the properties of the dc-SQUID potential in more detail. Then, we study the quantum
dynamics of the dc-SQUID which is equivalent to that of a fictitious particle trapped in
one of the wells of the potential. We will see under which conditions the dc-SQUID
behaves as a phase qubit. In this Section, the coupling between the SQUID and the ACPT
will be ignored such that we can consider the dc-SQUID as an independent element.
\subsection{dc-SQUID potential}
\begin{figure}
\centering
\includegraphics[width=0.8\linewidth]{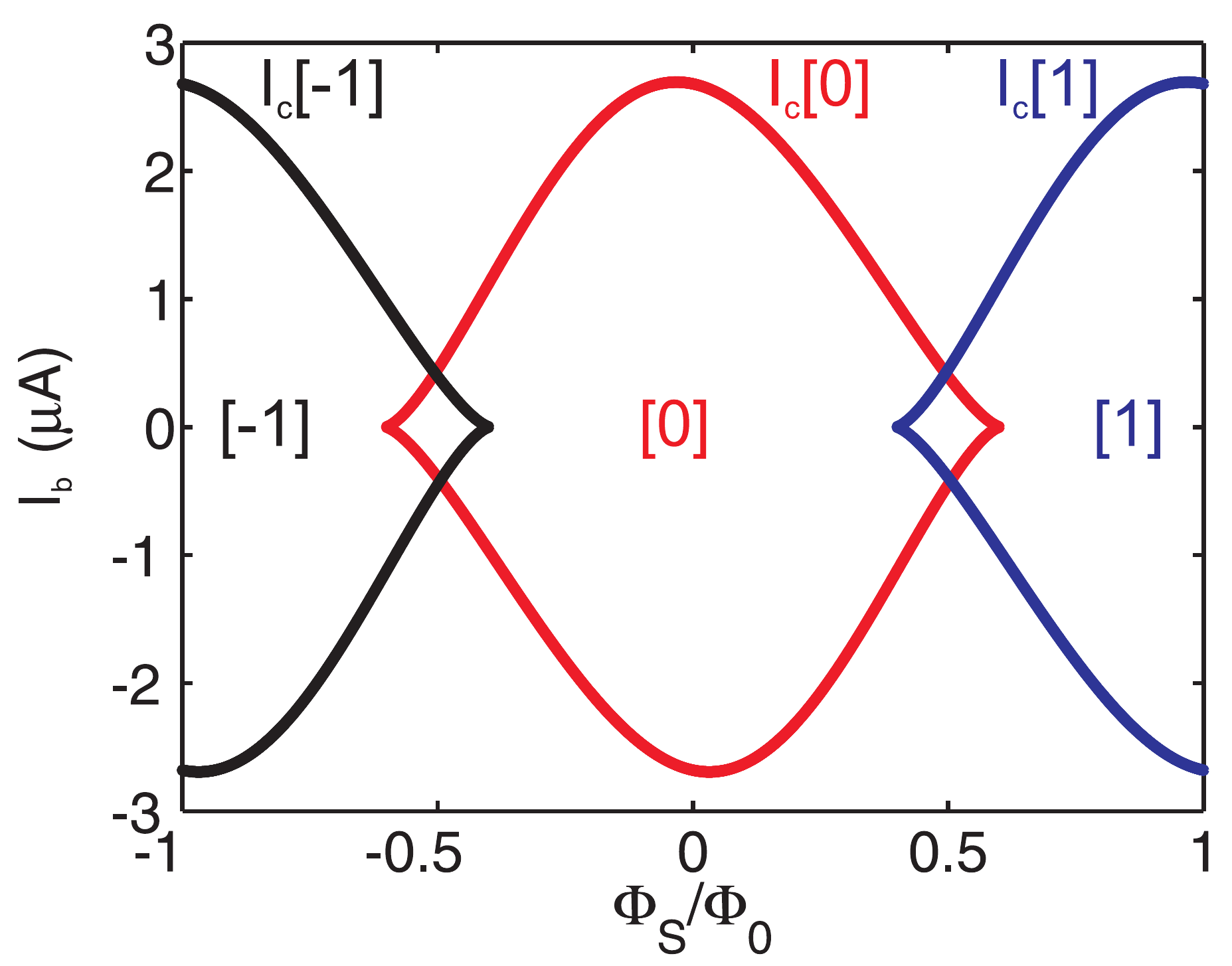}
\caption{Critical current $I_c[f]$ for well family [$f$] as a function of the flux
$\Phi_S$.} \label{fig:arches_SQUID}
\end{figure}

For an appropriate choice of bias parameters, i.e., the bias current $I_b$ and the flux
$\Phi_S$, the SQUID potential contains wells that can be regrouped in families
[$f$]~\cite{Lefevre_PRB92}. The index $f$ for the family [$f$] is related to the number
$f$ of flux quanta trapped in the dc-SQUID loop. The wells of the same family are located
along the direction $x$, periodically spaced by a distance $2\pi$, and separated by
saddle points. Since these wells have exactly the same geometry, the physical properties
of the SQUID are independent of the particular well in which the fictitious particle is
localized. Wells of the family [f] exist only if the bias current satisfies the relation
$I_c^-[f]<I_b<I_c^+[f]$, where $I_c^+[f]$ and $I_c^-[f]$ are the positive and negative
critical current of the family $[f]$, respectively. When $I_b=I_c^{\pm}[f]$, the local
minima of the wells of the family [$f$] and their closest saddle points coincide. The
critical currents depend strongly on the flux $\Phi_S$ as shown in
Fig.~\ref{fig:arches_SQUID} for the experimental parameters of the circuit studied in
Ref.~\cite{Fay2008}. Here, for a given flux, the absolute values of the critical currents
$I_c^+[f]$ and $I_c^-[f]$ are different. This difference originates from the finite
inductance asymmetry of the dc-SQUID ($\eta=0.28$) and disappears when $\eta=0$. For
almost any value of the flux, the potential is characterized by a unique family of wells,
except in the region close to $\pm\Phi_0/2$ where two families can coexist. This specific
region has been investigated recently in order to study the double escape path of the
particle, as well as to make the SQUID insensitive in first order to current fluctuations
\cite{Hoskinson2009}.
In the following we will discuss the dynamics of the dc-SQUID for the case where the
fictitious particle is trapped in one of the wells of the potential.
\subsection{Hamiltonian of the dc-SQUID}
In the full Hamiltonian (\ref{eq:Hamiltonien_total}) of the coupled circuit,  we isolate
the terms which only contain the operators $\widetilde{X}_\parallel$ and
$\widetilde{Y}_\perp$ and their conjugate momenta $\widetilde{P}_\parallel$ and
$\widetilde{P}_\perp$. These terms constitute the Hamiltonian of the dc-SQUID which takes
the form
\begin{eqnarray}
\widehat{H}_{S}&=&E_C^\parallel{\widetilde{{P}}_\parallel^2} +
\frac{1}{2}k_\parallel
{\widetilde{X}_\parallel^2}+\tilde{\sigma}_\parallel{\widetilde{{X}}_\parallel^3}\nonumber\\
&&+E_C^\perp{\widetilde{{P}}_\perp^2}+\frac{1}{2}k_\perp
{\widetilde{{Y}}_\perp^2}+\tilde{\sigma}_\perp
{\widetilde{{Y}}_\perp^3}\nonumber\\
&&+\beta_a  {\widetilde{{Y}}_\perp^2}{\widetilde{{X}}_\parallel} +
\beta_b {
\widetilde{{Y}}_\perp}{\widetilde{{X}}_\parallel^2}+\frac{(2e)^2}{C_{\parallel\perp}}{\widetilde{{P}}_\parallel}{\widetilde{{P}}_\perp},
\label{eq:Hamiltonien_SQUID_1}
\end{eqnarray}
with the charging energies $E_C^\parallel=(2e)^2/(2C_\parallel)$ and
$E_C^\perp=(2e)^2/(2C_\perp)$. The first three terms correspond to the Hamiltonian of a
fictitious particle of mass $m=\phi_0^2C_\parallel$ which is trapped in an anharmonic
potential along the longitudinal direction $X_\parallel$. These terms describe the
dynamics of an anharmonic oscillator of characteristic frequency
$\nu_p=\sqrt{k_\parallel/m}/(2\pi)$. The next three terms correspond to the Hamiltonian
of a fictitious particle of mass $m_\perp=\phi_0^2C_\perp$ which is trapped in an
anharmonic potential along the orthogonal direction $X_\perp$. These terms describe the
dynamics of an anharmonic oscillator of characteristic frequency
$\nu_\perp=\sqrt{k_\perp/m_\perp}/(2\pi)$. The anharmonicity of the two oscillators is
due to the cubic term which results from the non-linearity of the Josephson junction. The
three last terms of the dc-SQUID Hamiltonian mix the operators of the two oscillators and
consequently couple them. Finally, the two-dimensional dynamics of the SQUID is similar
to the dynamics of two coupled, one-dimensional oscillators. We proceed by introducing
the following dimensionless operators: $\widehat{X}_\parallel =
\sqrt{{h\nu_p}/{2E_C^\parallel}}\widetilde{X}_\parallel$, $\widehat{P}_\parallel =
-\sqrt{{2E_C^\parallel}/{h\nu_p}}\widetilde{P}_\parallel$, $\widehat{Y}_\perp =
\sqrt{{h\nu_\perp}/{2E_C^\perp}}\, \widetilde{Y}_\perp$ and $\widehat{P}_\perp =
-\sqrt{{2E_C^\perp}/{h\nu_\perp}}\, \widetilde{P}_\perp$, which verify the commutation
relations $[\widehat{X}_\parallel,\widehat{P}_\parallel]=i$ and
$[\widehat{X}_\perp,\widehat{P}_\perp ]=i$. With these new operators, the dc-SQUID
Hamiltonian can be rewritten as
\begin{eqnarray}
\widehat{H}_{S} &=&
\frac{1}{2}h\nu_p\left(\widehat{P}_\parallel^2+\widehat{X}_\parallel^2\right)-\sigma
h \nu_p
\widehat{X}_\parallel^3\nonumber\\
&&+\frac{1}{2}h\nu_\perp\left(\widehat{P}_\perp^2+\widehat{Y}_\perp^2\right)-\sigma_\perp
h
\nu_\perp \widehat{Y}_\perp^3\nonumber\\
&&+h\nu_a^* \widehat{Y}_\perp^2\widehat{X}_\parallel + h\nu_b^*
\widehat{Y}_\perp
\widehat{X}_\parallel^2+h\nu_c^*\widehat{P}_\perp\widehat
{P}_\parallel,\label{eq:Hamiltonien_SQUID_0}
\end{eqnarray}
where the parameters $\sigma$ and $\sigma_\perp$ correspond to the relative amplitude of
the cubic term compared to the quadratic term and hence are direct measure of the degree
of anharmonicity of the oscillators. The energies $h\nu_a^*$, $h\nu_b^*$ and $h\nu_c^*$
are the coupling energies between the two oscillators.

\begin{figure*}
\centering
\includegraphics[width=0.7\linewidth]{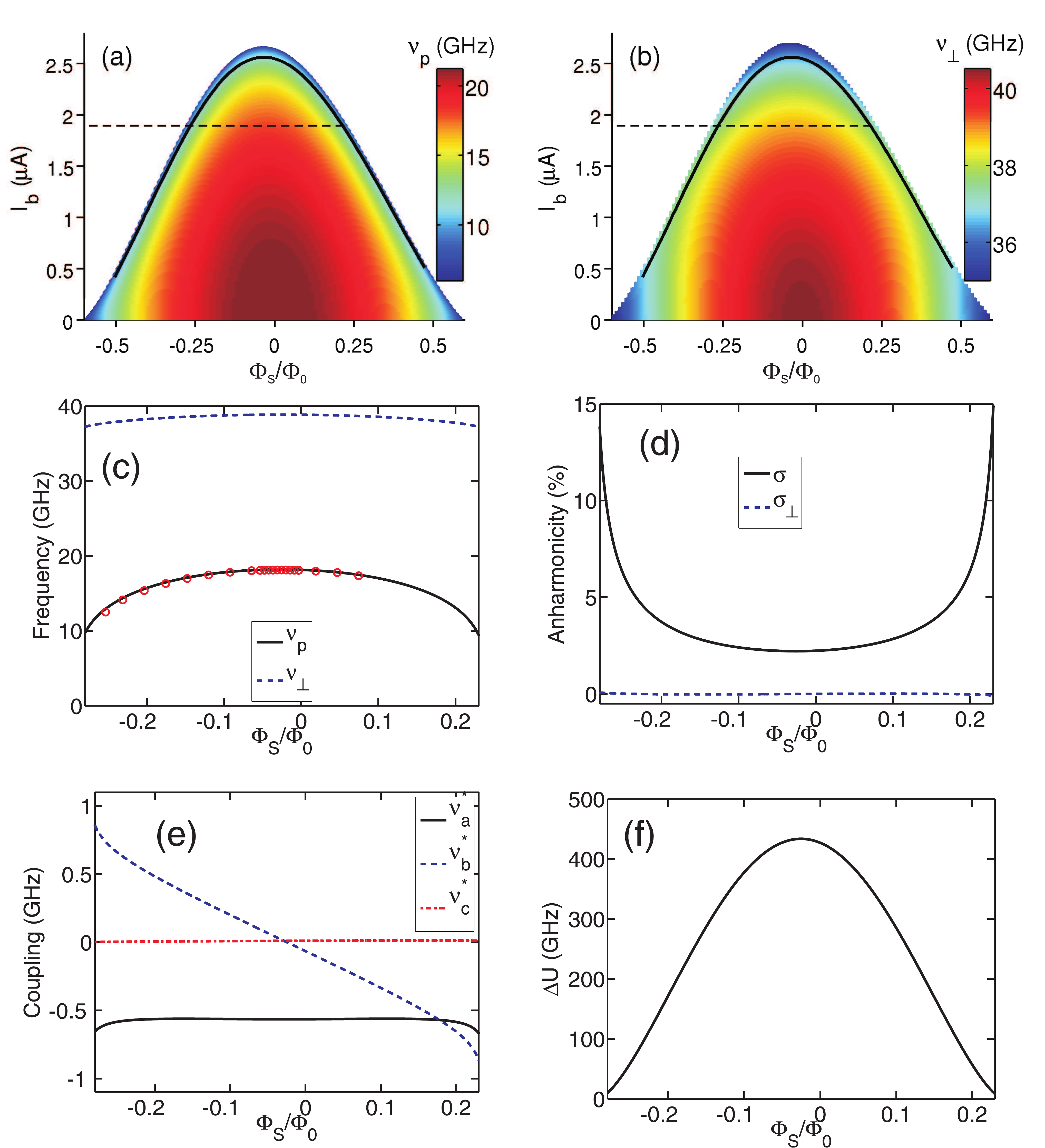}
\caption{ Numerical calculations of the parameters of the dc-SQUID Hamiltonian
(\ref{eq:Hamiltonien_SQUID_1}), using the experimental parameters of the circuit studied
in Ref.~\cite{Fay2008}. The fictitious particle associated to the SQUID is trapped in a
potential well of the family [0]. (a) Frequencies of the transverse and (b) orthogonal
oscillators, i.e. $\nu_p$ and $\nu_\perp$, respectively, as a function of the flux
$\Phi_S$ and the bias current $I_b$. The black line shows the measured switching current
of the dc-SQUID from Ref. \cite{FayThesis} which is close to the critical current.
(c)~Frequencies $\nu_p$ and $\nu_\perp$ as a function of $\Phi_S$ for $I_b=2$ $\mu$A,
i.e. for the bias points located on the dashed line of the Figs. (a) and (b). Measured
frequencies $\nu_S$ from Ref. \cite{FayThesis} are shown as red points. (d) Anharmonicity
parameters $\sigma$ and $\sigma_\perp$ of the longitudinal and orthogonal oscillators
respectively as a function of $\Phi_S$ for $I_b=2$~$\mu$A. (e) Different coupling
frequencies $\nu_a^*$, $\nu_b^*$ and $\nu_c^*$ between the two oscillators associated
with the terms in $\widehat{Y}_\perp^2\widehat{X}_\parallel$, $\widehat{Y}_\perp
\widehat{X}_\parallel^2$ and $\widehat{P}_\perp\widehat {P}_\parallel$ of $\widehat{H}_S$
(\ref{eq:Hamiltonien_SQUID_1}). (f) Energy $\Delta U$ for the potential barrier as a
function of $\phi_S$ for $I_b=2$~$\mu$A.} \label{fig:SQUID_parameters_draw}
\end{figure*}
\begin{table*}[htb!]
\centering
\begin{tabular}{c c|c c|c c c}
\hline\hline \multicolumn{2}{c|}{longitudinal oscillator} &
\multicolumn{2}{|c|}{transverse oscillator} &
\multicolumn{3}{|c}{coupling}
\\[0.5ex]
\hline $\nu_p$ & $\sigma$ & $\nu_\perp$ & $\sigma_\perp$ & $\nu_a^*$
& $\nu_b^*$ & $\nu_c^*$\\
$16.24$ GHz & $3.4$ \% & $38.26$ GHz & $-0.006$ \% & $-612$ MHz
& $-348$ MHz & $12$ MHz\\
\hline\hline
\end{tabular} \caption{Numerical values of the
parameters of the dc-SQUID Hamiltonian at the working point $\Phi_S=0.1\ \Phi_0$ and
$I_b=1.89\ \mu A$, calculated using the experimental parameters of the circuit studied in
Ref.~\cite{Fay2008}.} \label{tab:parametres_num_HS_verifie}
\end{table*}
Hereafter, we suppose that the particle is trapped in one of the wells of the family [0].
The geometry of this well varies as a function of the bias point $(I_b,\Phi_S)$ of the
circuit but does not change with the gate voltage $V_g$ and the flux $\Phi_T$. Therefore,
the different parameters of the dc-SQUID Hamiltonian only depend on $I_b$ and $\Phi_S$.
Fig.~\ref{fig:SQUID_parameters_draw} shows numerical calculations of the parameters of
the Hamiltonian under various biasing conditions, using the experimental parameters of
the circuit studied in Ref.~\cite{Fay2008} (see Appendix~\ref{appendix:parameters}).
Figs.~\ref{fig:SQUID_parameters_draw}(a) and \ref{fig:SQUID_parameters_draw}(b) show the
dependence of the transverse ($\nu_p$) and orthogonal ($\nu_\perp$) frequencies as a
function of the bias point. Generally, the frequency $\nu_\perp$ is always higher than 35
GHz and is at least twice higher than $\nu_p$. The frequency $\nu_p$  tends towards zero
when the bias point approaches the critical current line.
Figs.~\ref{fig:SQUID_parameters_draw}(c,d,e) show the dependence of the parameters of the
Hamiltonian as a function of the flux $\Phi_S$ for a fixed bias current of 1.89 $\mu$A.
Fig.~\ref{fig:SQUID_parameters_draw}(d) shows the two anharmonicity parameters of the two
oscillators. The anharmonicity $\sigma$ is typically around 3 \% and increases close to
the critical current line. The parameter $\sigma_\perp$ is very small regardless of the
bias, which leads to us to the conclusion that the orthogonal oscillator can be
considered as a harmonic one. Fig.~\ref{fig:SQUID_parameters_draw}(e) shows the different
coupling energies. The coupling frequency $\nu_c^*$ is of the order of 10 MHz and depends
only weakly on $\Phi_S$. The coupling frequencies $\nu_a^*$ and $\nu_b^*$ are generally
much higher than $\nu_c^*$ and depend on $\Phi_S$. Note that $\nu_b^*$ vanishes close to
$\Phi_S=0$ and $\nu_a^*$ is always negative. Numerical values of the parameters of the
Hamiltonian, for $\Phi_S=0.1$ $\Phi_0$ and $I_b=1.89$ $\mu$A, are given in
Tab.~\ref{tab:parametres_num_HS_verifie}. The energy $\Delta U$ of the potential barrier
(Fig.~\ref{fig:SQUID_parameters_draw}(f)) corresponds to the energy which separates the
local minimum of the well from its closest saddle point. This energy is not a parameter
of the dc-SQUID Hamiltonian (\ref{eq:Hamiltonien_SQUID_0}). Nevertheless, if $\Delta U$
is sufficiently small, typically on the order of $\nu_p$, the expression
(\ref{eq:Hamiltonien_SQUID_0}) of $\widehat H_S$ is too simplified. This occurs when the
bias point is close to the critical current line. One should then take into account the
coupling of the quantum levels inside the well to those outside the well. Note that this
coupling is responsible of the escape of the particle from the well
\cite{Balestro2003,Claudon2007}. This coupling will be neglected in the following,
assuming that the particle is always trapped in a sufficiently deep well.

\begin{figure}
\centering
\includegraphics[width=0.7\linewidth]{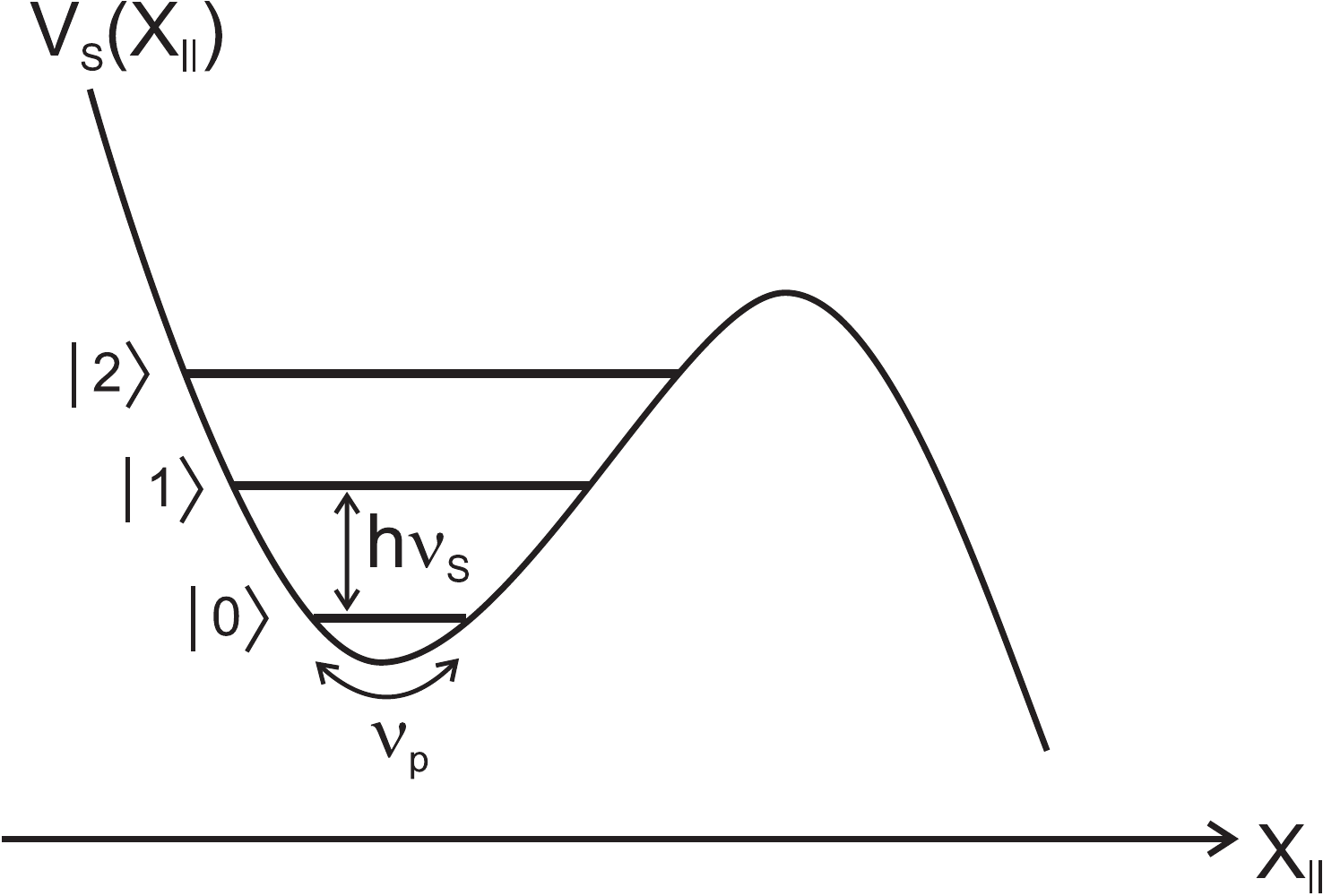}
\caption{Approximate one-dimensional potential of the dc-SQUID in the direction
$X_\parallel$. Eigenenergies of the first quantum states $|0\rangle$, $|1\rangle$ and
$|2\rangle$.} \label{fig:simplified_potential_SQUID}
\end{figure}

As $\nu_\perp\gg\nu_\parallel$, the quantum dynamics of the transverse oscillator is much
faster than that of the longitudinal oscillator. We will assume in the following that the
transverse oscillator is always in its ground state. It allows us to replace in $\widehat
H_S$ (\ref{eq:Hamiltonien_SQUID_0}) the operators of the transverse oscillator by their
average values given by $\langle\widehat{Y}_\perp \rangle=0$, $\langle\widehat{Y}_\perp^2
\rangle=1/2$ and $\langle\widehat{P}_\perp \rangle=0$. Only one of the three coupling
terms remains after this simplification. The coupling reads
$h\nu_a^*\widehat{X}_\parallel/2$ and can be seen as a modification of the bias current
$I_b$ of less than $0.5$ nA. This term will be neglected in the following. Under this
condition, the dynamics of the particle along the longitudinal direction is given by the
Hamiltonian
\begin{equation}
\widehat{
H}_{S}=\displaystyle{\frac{1}{2}}h\nu_p\left(\widehat{P}_\parallel^2+{\widehat{X}_\parallel}^2\right)-\sigma
h \nu_p {\widehat{X}_\parallel}^3. \label{eq:Hamiltonien_SQUID_2}
\end{equation}
In the following, the SQUID dynamics will be studied, using this simplified
Hamiltonian\cite{Claudon2007}. We denote  by $|n\rangle$ and $E_n$ the eigenstates and
the associated eigenenergies of $\widehat{H}_S$, respectively, such that
$\widehat{H}_S|n\rangle=E_n|n\rangle$, where $n$ is an integer number larger or equal to
zero. If the anharmonicity is weak ($\sigma\ll1$), the energies $E_n$ are given by a
straightforward perturbative calculation and we find
$E_n=(n+1/2)h\nu_p-15/4\sigma^2(n+1/2)^2h\nu_p$.
Fig.~\ref{fig:simplified_potential_SQUID} shows the approximate potential of the dc-SQUID
and the three first eigenenergies. When the anharmonicity of the dc-SQUID is sufficiently
large, the dynamics of the dc-SQUID in the presence of an external microwave perturbation
involves only the two first levels $|0\rangle$ and $|1\rangle$ \cite{Claudon2004}. In
that case, the dc-SQUID behaves as a qubit. Since for the dc-SQUID the Josephson energy
is much larger than the charging energy, the fluctuations of the phase
$\widetilde{X}_\parallel$ are much smaller than those of the charge
$\widetilde{P}_\parallel$. For this reason, the dc-SQUID is referred to as a phase qubit
\cite{Hoskinson2009}. The dc-SQUID Hamiltonian can be rewritten in the basis
$(|1\rangle,|0\rangle)$, using the Pauli matrices (See
Appendix~\ref{annexe:Pauli_matrix}), as $\widehat{H}_S=h\nu_s/2\sigma_z^S$ where
$\nu_S\equiv(E_1-E_0)/h$ is the characteristic qubit frequency. The frequency $\nu_S$ can
be approximated in first order with respect to the anharmonicity $\sigma$ as
$h\nu_S=h\nu_p(1-(15/2)\sigma^2)$. We see that the frequency $\nu_S$ equals the plasma
frequency $\nu_p$ if the dc-SQUID anharmonicity is zero and decreases with increasing
anharmonicity (see Fig.~\ref{fig:SQUID_parameters_draw}(c)). When the anharmonicity
$\sigma$ is zero, the dc-SQUID behaves as a harmonic oscillator described by the typical
Hamiltonian $\hat{H_S}=h\nu_p(\hat a \hat a^\dag+1/2)$, where $\hat a^{\dag}$ and $\hat
a$ are the one-plasmon creation and annihilation operators, respectively.

\section{Asymmetric Cooper Pair Transistor (ACPT)}

The dc-SQUID Hamiltonian is only one of the terms that appear in the full Hamiltonian of
the circuit. It is coupled to a second term associated with the asymmetric Cooper Pair
transistor (ACTP). This section is dedicated to the theoretical analysis of the ACPT
dynamics, neglecting its coupling with the dc-SQUID. We will first build and analyze the
ACPT Hamiltonian and will show that the ACPT can be viewed as a charge qubit. We will
then describe the ACPT by using only two charge states. The errors on the level of the
eigenenergies and the eigenstates induced by this simplified description will be
estimated.

\subsection{ACPT Hamiltonian}
\begin{figure}
\centering
\includegraphics[width=0.7\linewidth]{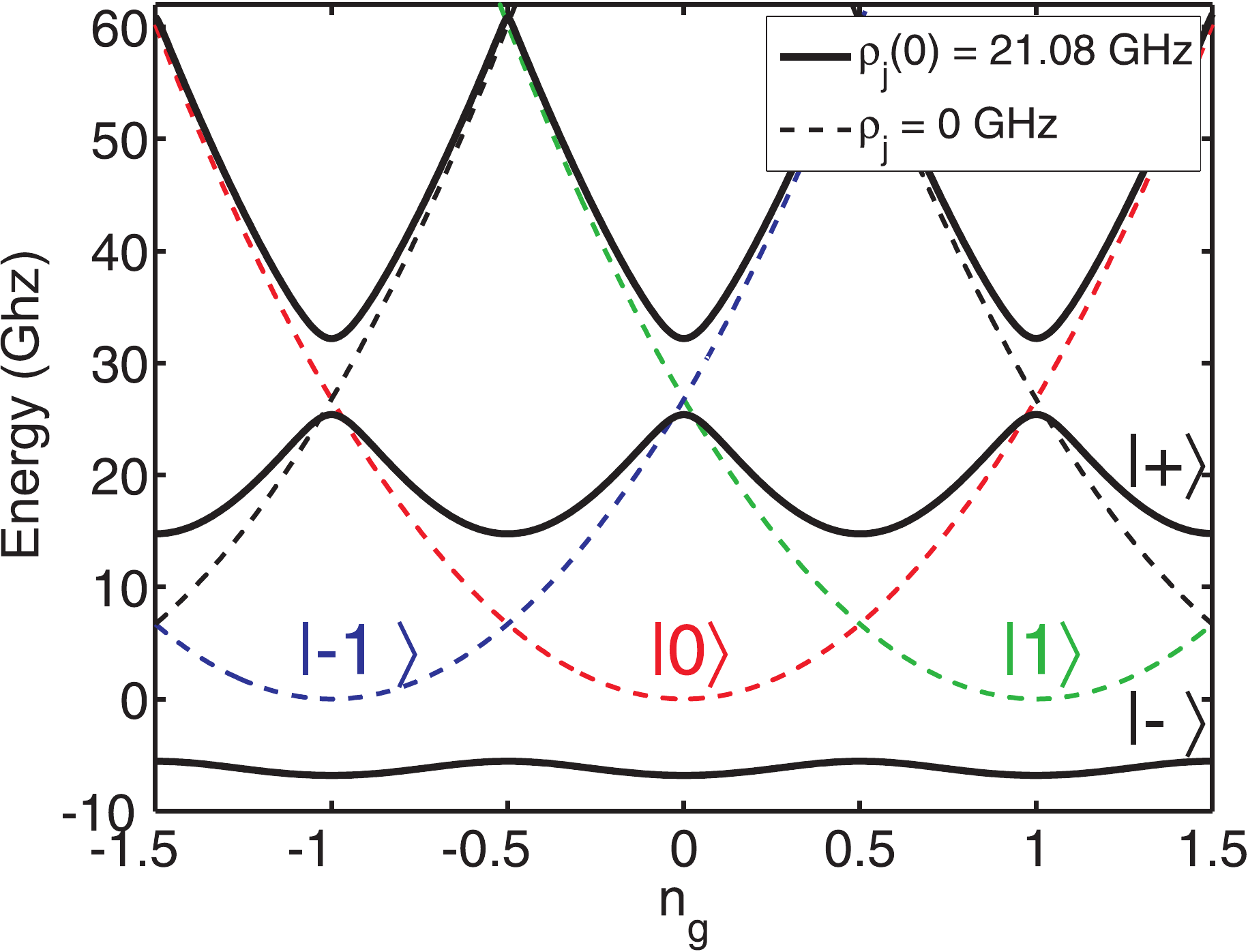}
\caption{Energy spectrum of the ACPT as a  unction of $n_g$ in the case of a zero
Josephson coupling (dashed lines) and in the case of $\rho_j(\delta=0)=21.08$ GHz (full
lines). These energy spectra have been calculated numerically, using the parameters from
Ref.~\cite{Fay2008}: $E_J^T=21.08$ GHz, $E_C^T=26.76$ GHz and $\mu=-41.6 \%$. The states
$|-\rangle$ and $|+\rangle$ associated to the two lowest energy bands correspond to the
states of the charge qubit.} \label{fig:ACPT_spectrum}
\end{figure}

\begin{figure}
\centering
\includegraphics[width=0.7\linewidth]{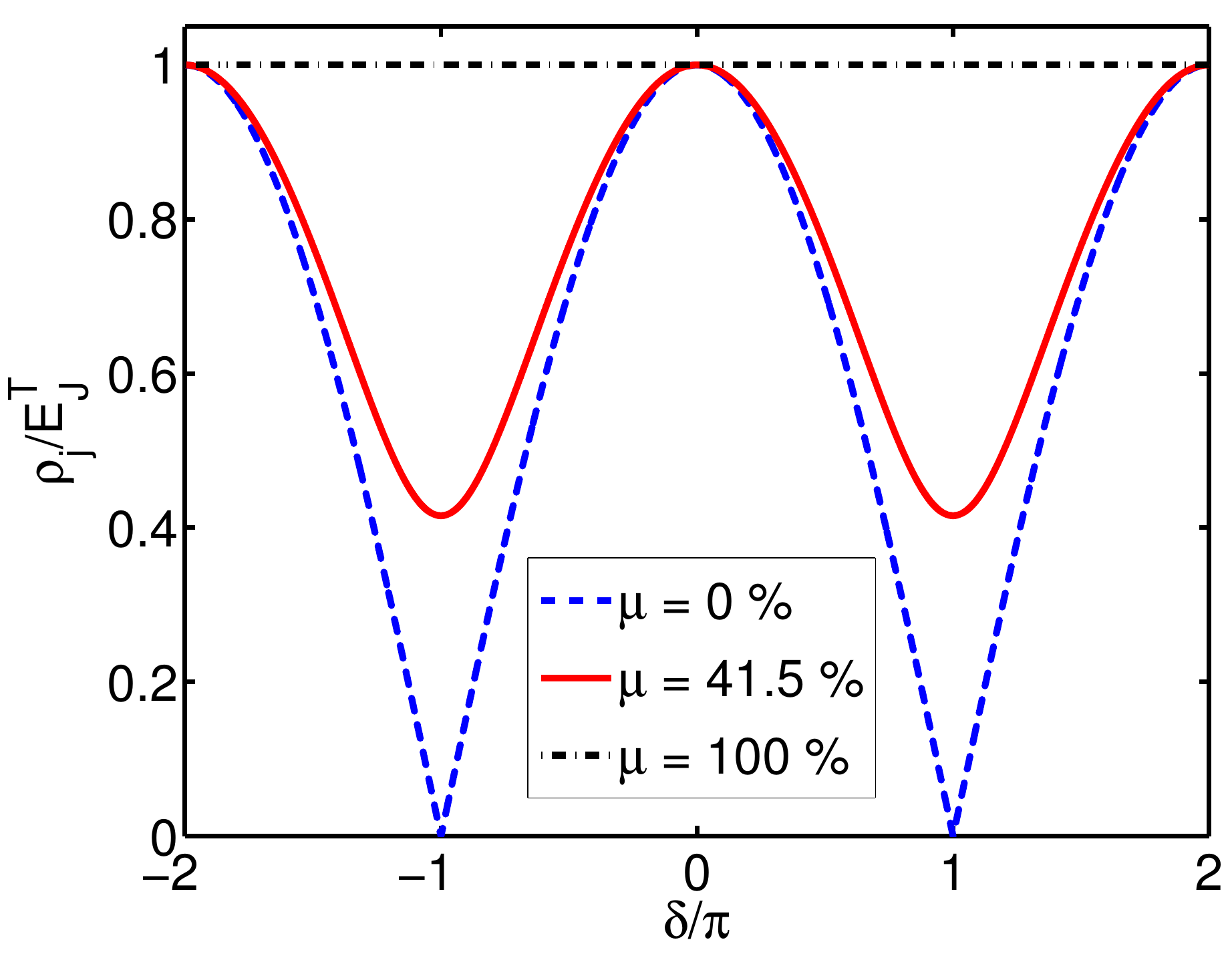}
\caption{Josephson coupling $\rho_j$ as a function of the superconducting phase $\delta$
across the ACPT, for three different Josephson asymmeties: $\mu=0$ for a symmetric
transistor, $\mu=-41.5 \%$ for an asymmetric transistor and $\mu=100\%$ for a Cooper pair
box.} \label{fig:rhoj}
\end{figure}

\begin{figure}
\centering
\includegraphics[width=0.7\linewidth]{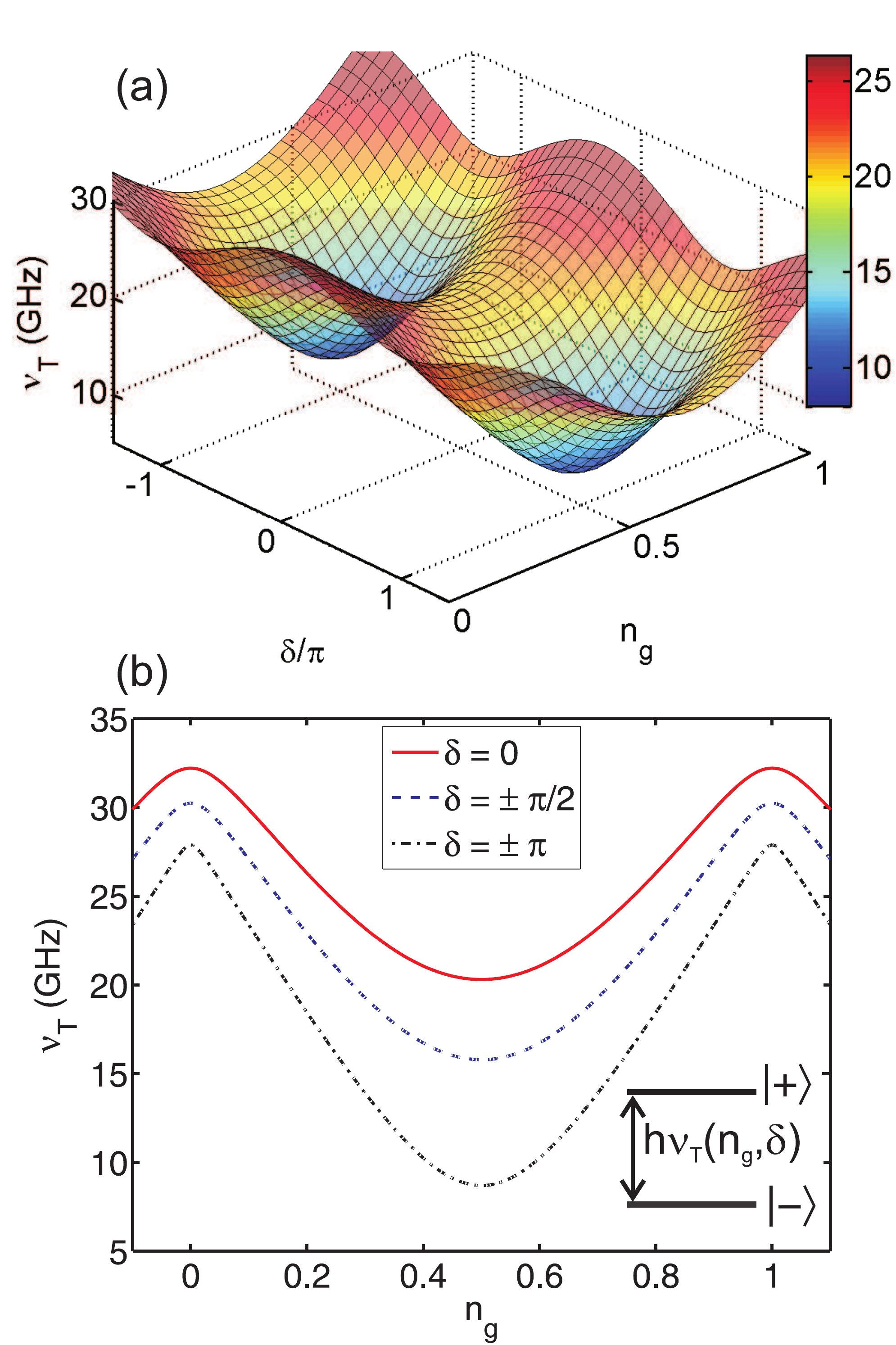}
\caption{(Color online). (a) Frequency $\nu_T$ of the charge qubit as a function of $n_g$
and $\delta$ for the experimental parameters of Ref.~\cite{Fay2008}: $E_C^T=1.28\ k_B$K,
$E_J^T=1.01\ k_B$K and $\mu=-41.6 \%$. (b) Frequency $\nu_T$ as a function of $n_g$ for
$\delta=0$, $\delta=\pm\pi/2$ and $\delta=\pi$.} \label{fig:energies_acpt}
\end{figure}

\begin{figure}
\centering
\includegraphics[width=0.7\linewidth]{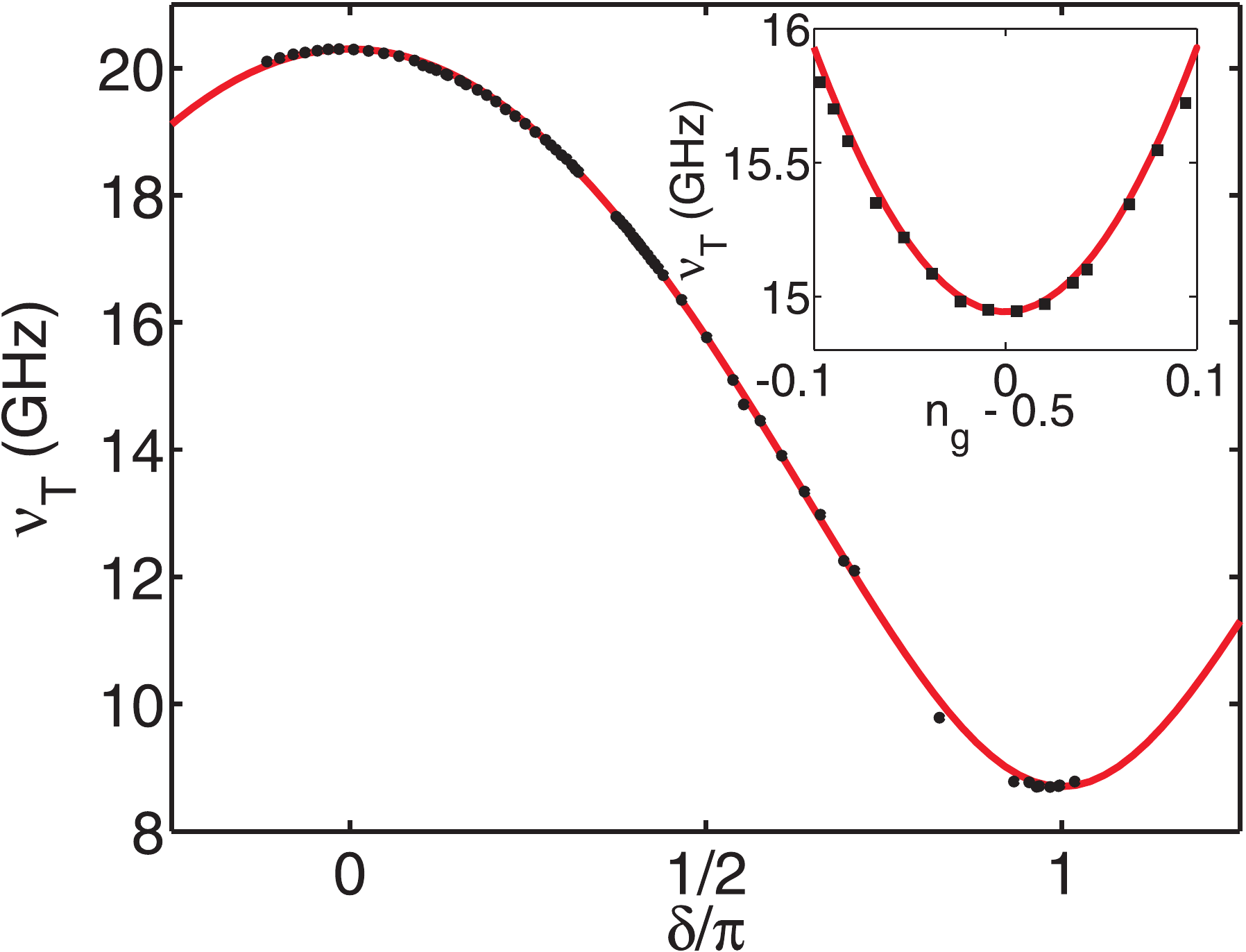}
\caption{(Color online). Frequency $\nu_T$ as a function of the phase $\delta$. The
points are the experimental data from Ref.~\cite{Fay2008}. The solid curve is the
theoretical frequency $\nu_T$ calculated by diagonalizing the ACPT Hamiltonian
(\ref{eq:HACPT_chargerepresentation}) with the parameters $E_J^T=21.08$ GHz,
$E_C^T=26.76$ GHz and $\mu=-41.6$ $\%$.  Inset: Frequency $\nu_T$ as a function of $n_g$
for $\delta=0.549\pi$. The points are the experimental frequencies $\nu_T$ and the solid
line is the theoretical frequency $\nu_T$ calculated as before.} \label{fig:wt_vs_delta}
\end{figure}

The ACPT Hamiltonian is identified by isolating from the full Hamiltonian of the circuit
(\ref{eq:Hamiltonien_total}) the terms which only contain the operators $\hat\psi$ and
$\hat n$. After some straightforward algebra, the ACPT Hamiltonian reads
\begin{eqnarray}
\hat H_{ACPT}=&&E_C^T\left(\hat n -
n_g\right)^2-E_J^T\cos(\delta/2)\cos(\hat\psi-\delta/2)\nonumber\\&&+\mu\sin(\delta/2)\sin(\hat
\psi - \delta/2)), \label{eq:Hamiltonien_ACPT_1}
\end{eqnarray}
where $E_J^T\equiv E_{J1}^T+E_{J2}^T$ and $E_C^T\equiv (2e)^2/(2C_T)$ are the Josephson
and charging energies of the ACPT, respectively. It is the generalization of the
Quantronium Hamiltonian\cite{Vion_Science02} for which asymmetries in critical current
and capacitance of the Cooper pair transistor were neglected. The gate-charge $n_g\equiv
C_gV_g/(2e)$ corresponds to the number of Cooper pairs induced by the voltage applied to
the gate-capacitance. The charge states $|n_{2e}\rangle$ are the eigenvectors of the
charge operator $\hat n$, such $\hat n|n_{2e}\rangle=n|n_{2e}\rangle$ where
$n=0,\pm1,\pm2...$ is the number of excess Cooper pairs on the transistor island. Using
the commutation relation $[\hat \psi, \hat n]=i$, we identify the action of the operator
$e^{i\hat\psi}=\sum_n|n_{2e}\rangle\langle (n+1)_{2e}|$ which decreases the number of
Cooper pairs on the island by one unit. In the charge representation, the ACPT
Hamiltonian can be written as
\begin{eqnarray}
\hat H_{ACPT}=&& E_C^T(\hat n - n_g)^2  -
\displaystyle{\frac{\rho_j(\delta)}{2}}\bigg[ \sum \limits_n
e^{-i(\delta/2+\chi)} |n_{2e}\rangle\langle (n+1)_{2e}| \nonumber\\&&+ \sum \limits_n
e^{i(\delta/2+\chi)}|(n+1)_{2e}\rangle\langle n_{2e}|\bigg],
\label{eq:HACPT_chargerepresentation}
\end{eqnarray}
with ${E_J^T}/{2}( \cos (\delta/2)+ i \mu \sin (\delta/2) ) \equiv \rho_j(\delta)/2
e^{i\chi}$, $\tan \chi = \mu\tan(\delta/2)$ and $
\rho_j(\delta)^2={E_J^T}^2(\cos^2(\delta/2)+\mu^2\sin^2(\delta/2))$. The ACPT Hamiltonian
is composed of a charging and a Josephson term which are proportional to $E_C^T$ and
$\rho_j(\delta)$, respectively. Let us focus first on the case of a zero Josephson
coupling ($\rho_j=0$). The eigenstates of the ACPT Hamiltonian are then the charge states
$|n_{2e}\rangle$ with the associated eigenenergies $E_C^T(n-n_g)^2$.
Fig.~\ref{fig:ACPT_spectrum} shows the energy spectrum of the ACPT as a function of $n_g$
for a charging energy $E_C^T=1.28$ K. This spectrum consists of a series of parabolas,
each parabola being associated with a specific charge state $|n_{2e}\rangle$  with a
minimum energy for $n_g=n$. Notice that when the energy difference between the ground
charge state and the first excited charge state is much larger than $k_BT$, the charge on
the island is well quantized leading to the Coulomb blockade phenomena \cite{Averin1991}.
For $n_g=0.5$, the energy parabolas of the states $|0_{2e}\rangle$
and $|1_{2e}\rangle$ cross each other and the states $|0_{2e}\rangle$ and
$|1_{2e}\rangle$ are degenerate. This degeneracy is lifted by the Josephson term which
couples the neighboring charge states to each other. The amplitude of this Josephson
coupling is given by $\rho_j$. Fig.~\ref{fig:rhoj} shows the dependence of $\rho_j$ on
$\delta$ for three different Josephson asymmetries. Since $\rho_j$ is $2\pi$ periodic in
$\delta$, the range of $\delta$ has been restricted to the interval between $-2\pi$ and
$2\pi$. In the case of a symmetric transistor ($\mu=0$), the Josephson coupling is
maximum for $\delta=0$, equal to $E_J^T$; it is zero at $\delta=\pm\pi$. For a finite
asymmetry $\mu$, the Josephson coupling reaches a maximum for $\delta=0$ and equals
$E_J^T$, and a minimum equals to $\mu E_J^T$ for $\delta=\pm\pi$. For a Cooper pair box
($\mu=\pm100\%$), the Josephson coupling does not depend on $\delta$ and remains equal to
$E_J^T$. The Josephson coupling then depends strongly on the Josephson asymmetry,
especially for $\delta=\pm\pi$ where it can vary from zero to $E_J^T$.
%
The full energy spectrum of the ACPT, which takes into account the Josephson coupling, is
calculated numerically by diagonalizing the Hamiltonian
(\ref{eq:HACPT_chargerepresentation}) with 8 charge states. It is plotted as a function
of $n_g$ in Fig.~\ref{fig:ACPT_spectrum} for a fixed superconducting phase $\delta=0$ and
for the parameters of \cite{Fay2008}: $E_C^T=1.28 K$, $E_J^T=1.01 k_BK$ and $\mu=-41.6
\%$. The spectrum consists of several energy bands which do not cross each other, leading
to an energy gap between the lowest and the first bands. We associate with these two
bands the ground state $|-\rangle$ and the first excited state $|+\rangle$, respectively.
These two states correspond to the states of a qubit, with a characteristic energy
$h\nu_T$. Note that to operate the ACPT as a qubit, the gate-charge should be close to
0.5 (modulo 1) where the transition energy from the state $|+\rangle$ to the third level
is higher than the frequency $\nu_T$. Indeed, when the ACPT Hamiltonian
(\ref{eq:HACPT_chargerepresentation}) is subject to an adequate perturbation, the quantum
dynamics of the ACPT will only involve the states $|-\rangle$ and $|+\rangle$. Because
$E_J^T>E_C^T$ the states $|-\rangle$ and $|+\rangle$ can be written as a superposition of
a few charge states (typically four). For this reason, the qubit formed by the ACPT is
referred to as a charge qubit. The Hamiltonian of the charge qubit can be written using
the Pauli matrix (see appendix \ref{annexe:Pauli_matrix}) as
\begin{equation}
 \frac{h\nu_T}{2}\hat\sigma_z^T.
\end{equation}
The frequency $\nu_T$ of the charge qubit depends on the two parameters $\delta$ and
$n_g$ as shown in Fig.~\ref{fig:energies_acpt}(a). This frequency, as is $\rho_j$, is
also $2\pi$ periodic as a function of $\delta$. It is maximum and minimum at the points
($\delta=0$,$n_g=1/2$) and ($\delta=\pi$,$n_g=1/2$), respectively. These are optimal
points where the ACPT is, in first order, insensitive  to charge, flux and current
fluctuations. Fig.~\ref{fig:wt_vs_delta} shows the experimental frequency of the charge
qubit measured in Ref.~\cite{Fay2008} as a function of $\delta$. The theoretical fit,
shown in red, is obtained by diagonalizing the ACPT Hamiltonian
(\ref{eq:HACPT_chargerepresentation}) in a basis of 8 charge states. It allows us to
accurately find the parameters of the charge qubit, i.e. the Josephson energy
$E_J^T=21.08$ GHz, the charging energy $E_C^T=26.76$ GHz and the Josephson asymmetry
$\mu=-41.6$~$\%$. Nevertheless, the capacitance asymmetry can not be extracted from the
fit since it does not enter in the ACPT Hamiltonian
(\ref{eq:HACPT_chargerepresentation}). The Josephson energy and the capacitance of a
Josephson junction of the ACPT are both proportional to the junction surface and,
therefore, the capacitance and Josephson asymmetries are equal in first approximation. We
will see in Sec.~\ref{sec:coupling} that the capacitance asymmetry can be extracted from
the coupling between the dc-SQUID and the ACPT and we will find  $\lambda=0.875\mu$.

\subsection{Description of the ACPT with two charge levels}
\label{sec:ACPT_two_levels} If the Josephson coupling $\rho_j$ is much smaller than the
charging energy $E_C^T$ and if $n_g\approx0.5$, the qubit states can be expressed as a
superposition of the two charge states $|0_{2e}\rangle$ and $|1_{2e}\rangle$. In that
case, the Hamiltonian of the transistor is simply given by its matrix form $\hat
H_{ACPT}^0$ which reads, in the charge basis ($|0_{2e}\rangle$,$|1_{2e}\rangle$),
\begin{equation}
\hat H_{ACPT}^0 = \left(
                              \begin{array}{cc}
                                E_C^T n_g^2 & -\rho_j e^{-i(\delta/2+\chi)}/2 \\
                                -\rho_j e^{i(\delta/2+\chi)}/2 & E_C^T(1-n_g)^2 \\
                              \end{array}
                            \right).
\label{eq:Hacpt0}
\end{equation}
The eigenvalues of this simplified Hamiltonian are given by $E_\pm^0 = \frac{1}{2}
E_C^T\left(n_g^2 + (1-n_g)^2\right) \pm \frac{1}{2} \sqrt{{E_C^T}^2(1-2 n_g)^2 +
\rho_j^2}$, and the qubit energy reads $h\nu_T^0=E_+^0-E_-^0$. Note that for $n_g=0.5$,
we have $h\nu_T^0=\rho_j$. The eigenstates $|-_0\rangle$ and $|+_0\rangle$, associated
with the energies $E_-$ and $E_+$, can be written as a function of the charge states as
\begin{equation}
\left\{ \begin{aligned}
&|+_0\rangle = \alpha^* |0_{2e}\rangle + \beta
|1_{2e}\rangle\\
&|-_0\rangle = -\beta^* |0_{2e}
\rangle + \alpha |1_{2e}\rangle
\end{aligned} \right.
\label{eq:charge_states}
\end{equation}
with $\alpha=\cos(\theta/2) e^{i(\delta/2+\chi-\pi)/2}$, $\beta=\sin(\theta/2)
e^{i(\delta/2+\chi-\pi)/2}$ and $\tan \theta = -\frac{2|\rho_j|}{E_C^T (1-2 n_g)}$ ($\theta\in[0,\pi]$).

\begin{figure}
\centering
\includegraphics[width=0.7\linewidth]{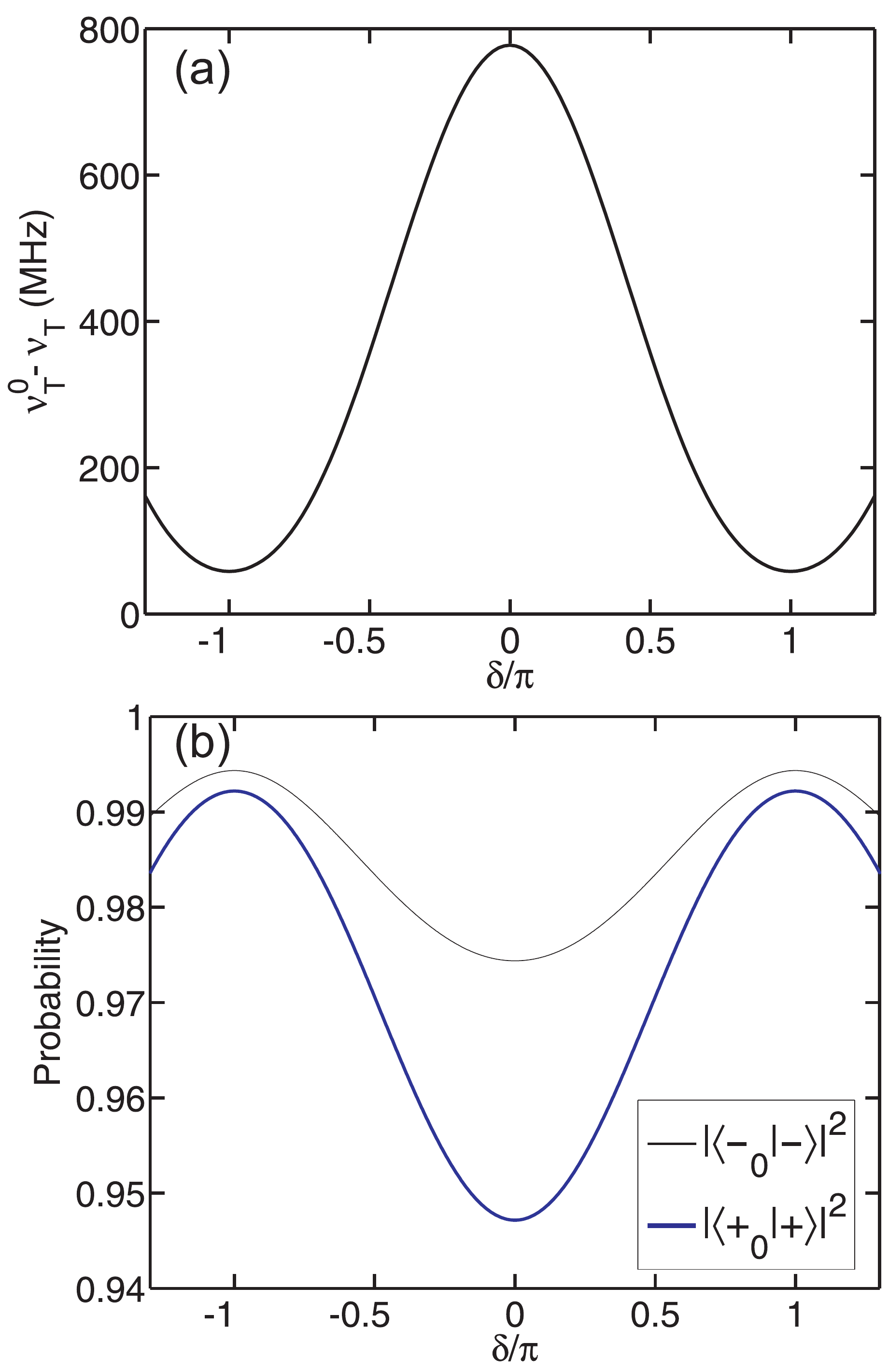}
\caption{(Color online). Errors in the energies and states of the charge qubit of
Ref.~\cite{Fay2008} induced by describing it with only two charge states. (a) Difference
between the qubit frequency $\nu_T^0$ calculated with two charge states and its real
frequency $\nu_T$ as fonction of $\delta$ for $n_g=0.5$. (b) Probabilities
$|\langle-_0|-\rangle|^2$ (black) and $|\langle+_0|+\rangle|^2$ (blue) as a function of
$\delta$ for $n_g=0.5$.} \label{fig:errors_acpt}
\end{figure}

For the ACPT studied experimentally in Ref.~\cite{Fay2008}, the condition
$\rho_j(\delta)\ll E_C^T$ is too strong, especially at $\delta=0$ where the ratio
$\rho_j/E_C^T\approx79 \%$ is maximum. For this reason, we would like to quantify the
error induced by the description of the ACPT with only two charge states. Let us first
focus on the error in the qubit energy. Fig.~\ref{fig:errors_acpt} shows the difference
between the frequencies $\nu_T^0$ calculated with two charge states and the real
frequency $\nu_T$ as a function of $\delta$ for $n_g=0.5$. This difference is minimum for
$\delta=\pm\pi$ and equals $58$ MHz, corresponding to an error in energy of $0.5\%$. For
$\delta=0$, the difference is maximum and equals $780$ MHz, corresponding to an error in
energy of $3.8\%$.

We now discuss the error made on the level of the states $|-_0\rangle$ and $|+_0\rangle$.
The non-reduced Hamiltonian of the ACPT can be rewritten as a function of
$\widehat{H}^0_{ACPT}$ as $\widehat{H}_{ACPT}=\widehat{H}^0_{ACPT}+\widehat{W}$, where
$\widehat{W}$ is a perturbative term which takes the form $\widehat{W}=-
\frac{\rho_j(\delta)}{2}\sum \limits_{n\neq0}\left[ e^{-i(\delta/2+\chi)}
|n\rangle\langle n+1| + e^{i(\delta/2+\chi)}|n+1\rangle\langle n|\right] + E_C^T\sum
\limits_{n\neq\{0,1\}}(n-n_g)^2|n\rangle\langle n|$. Using first-order perturbation
theory, we calculate the probability $|\langle-_0|-\rangle|^2$
($|\langle+_0|+\rangle|^2$) that the state $|-\rangle$ ($|+\rangle$) is in the the state
$|-_0\rangle$ ($|+_0\rangle$). Fig.~\ref{fig:errors_acpt} shows the dependence of these
probabilities as a function of $\delta$ for $n_g=0.5$. The probability
$|\langle-_0|-\rangle|^2$ is always smaller than the probability
$|\langle+_0|+\rangle|^2$. Indeed, the energy of the state $|+_0\rangle$ is closer to the
charging energies of the states $|-1_{2e}\rangle$ and $|2_{2e}\rangle$ and consequently
more perturbed by these two states. The probabilities are minimum for $\delta=0$
($|\langle-_0|-\rangle|^2=97.4\%$ and $|\langle-_0|-\rangle|^2=94.7\%$) and, therefore,
for this value of the phase, the error made by considering the states of the ACPT as
$|-_0\rangle$ and $|+_0\rangle$ is maximum. Fig.~\ref{fig:errors_acpt} shows the
dependence of the probabilities $|\langle-_0|-\rangle|^2$ and $|\langle+_0|+\rangle|^2$
as a function of the ratio $E_J^T/E_C^T$ for $\delta=0$. Theses probabilities have been
calculated with a first-order perturbation theory and also numerically by using 8 charge
states. We see that when $E_J^T/E_C^T<1$, the first-order perturbation theory agrees
quite well with the numerical simulations. But when $E_J^T/E_C^T>1$, the analytical
calculation gives probabilities significantly lower than the numerical one. The
probabilities $|\langle-_0|-\rangle|^2$ and $|\langle+_0|+\rangle|^2$ decrease with
increasing $E_J^T/E_C^T$, which is explained by the fact that the Josephson coupling
mixes more and more the charge states $|0_{2e}\rangle$ and $|1_{2e}\rangle$ with the
other, closer-in-energy charge states. In order to simplify the calculation of the
analytical expression of the coupling (see below), we consider in the following that
$|-\rangle=|-_0\rangle$ and $|+\rangle=|+_0\rangle$, but do not approximate the qubit
frequency to $\nu_T^0$.

\begin{figure}
\centering
\includegraphics[width=0.7\linewidth]{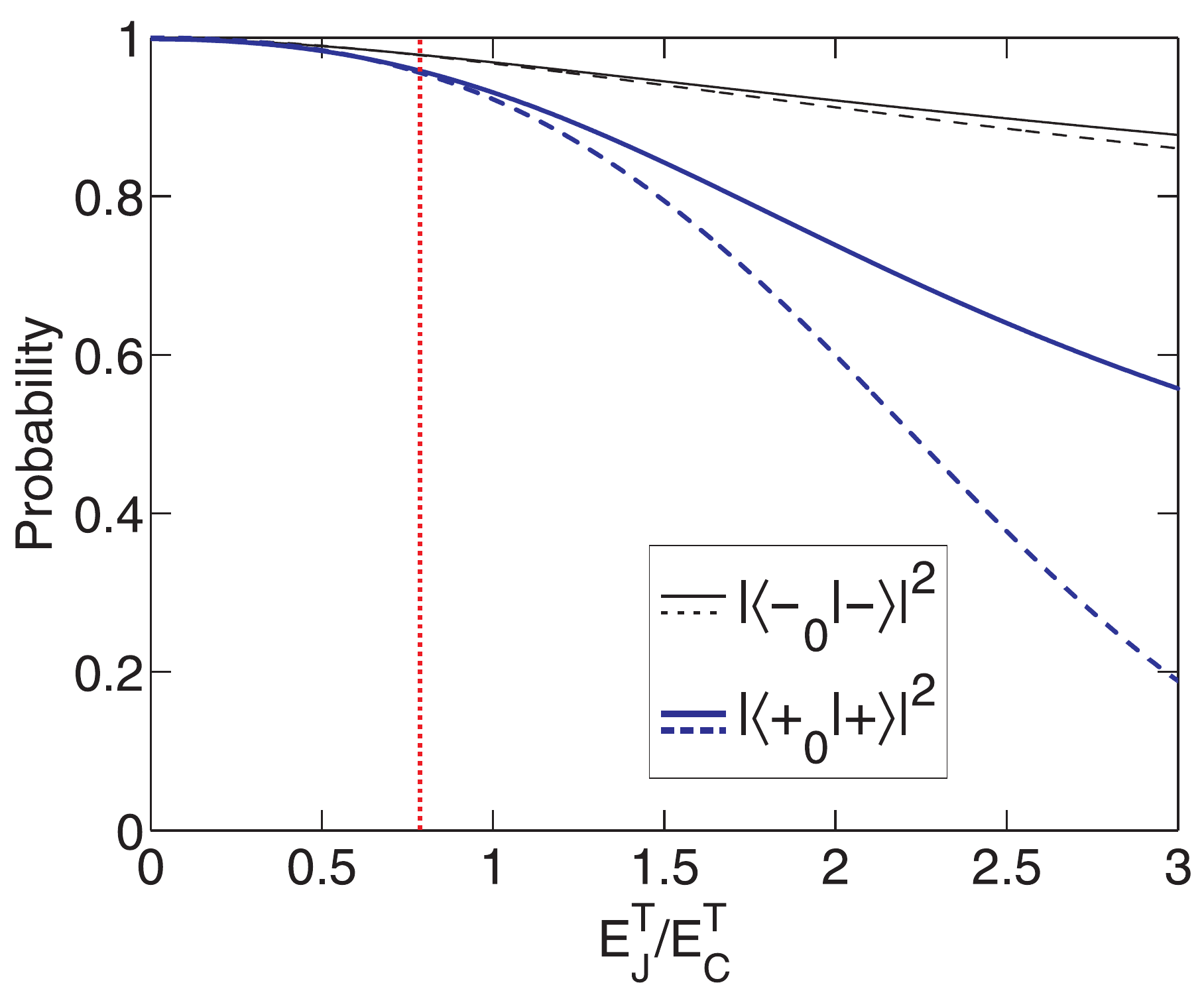}
\caption{(Color online). Probabilities $|\langle-_0|-\rangle|^2$ (black) and
$|\langle+_0|+\rangle|^2$ (blue) as a function of the ratio $E_J^T/E_C^T$ for $n_g=0.5$
and $\delta=0$. Numerical calculations of these probabilities (full lines) are compared
with first-order perturbation calculations (dashed lines). The vertical dashed line
indicates the values of these probabilities at the ratio $E_J^T/E_C^T=79\%$ corresponding
to those of the ACPT studied in Ref.~\cite{Fay2008}. } \label{fig:errors_ratio_acpt}
\end{figure}

\section{Coupling}
\label{sec:coupling}
So far, we have considered independently the quantum dynamics of the longitudinal mode of the dc-SQUID and the ACPT. However, in the studied circuit the dc-SQUID and the ACPT are connected in parallel and therefore coupled to each other. The independent dynamics of the ACPT and the dc-SQUID has to be reconsidered especially when the two qubits are close to resonance ($\nu_S\approx\nu_T$). In this case the coupling effects are the strongest. In this section, we derive the expression of the coupling Hamiltonian by considering the ACPT as a charge qubit and the dc-SQUID as either a tunable harmonic oscillator or a phase qubit. We will see that the total coupling is the sum of two distinct contributions: a capacitive and an inductive Josephson coupling.

\subsection{Capacitive coupling}
The capacitive coupling Hamiltonian couples by definition the charge of the dc-SQUID with that of the ACPT. It reads
\begin{equation}
\hat
H_{Coupl,Capa}=-\frac{(2e)^2}{C_{n\parallel}}\sqrt{\frac{h\nu_p}{2E_C^\parallel}}\widehat{P}_\parallel
(\hat n -n_g).\label{eq:H_coup,capa}
\end{equation}
%
%
We consider hereafter two different limits for the dc-SQUID in order to simplify this capacitive coupling Hamiltonian.

The first limit corresponds to a dc-SQUID with an anharmonicity factor $\sigma$ equal to
zero. This limit can be achieved when the fictitious particle associated with the
dc-SQUID is trapped in a deep well, which is generally true when the dc-SQUID is biased
at the working point $I_b=0$ and $\Phi_S=0$. Under this condition, the dc-SQUID behaves
as a harmonic oscillator and several levels are involved in the dynamics. The
momentum operator in the charge coupling Hamiltonian is then given by
$\widehat{P}_\parallel=(\hat a^\dag+\hat a)/\sqrt{2}$, where $\hat a^\dag$ and $\hat a$
are the one-plasmon creation and annihilation operators. If we describe the charge
qubit with two charge states (Eq.~(\ref{eq:charge_states})), the
operator $\hat n$ can be written as $\widehat{n}=\frac{1}{2}\,\hat
I+\frac{1}{2}\sin(\theta)\hat\sigma_x^T-\frac{1}{2}\cos(\theta)\hat\sigma_z^T$, where
$\hat I$ is the identity operator and $\sigma_x^T$ and $\sigma_y^T$ are the Pauli
matrices defined in the eigenstate basis of the charge qubit. For $n_g=1/2$, we have
$\theta=\pi/2$ and the charge coupling Hamiltonian simplifies to
\begin{equation}
\widehat{H}_{Coupl,Capa}=-E_{c,c}\,(\hat a^\dag+\hat a)\hat{\sigma}_x^T/\sqrt{2},
\end{equation}
where $E_{c,c}=e^2/C_{n\parallel}\sqrt{h\nu_p/E_C^\parallel}$ determines the strength of the capacitive coupling.

The second limit is realized when the anhamonicity $\sigma$ is typically around 3 $\%$ or larger; then the dc-SQUID can be described as a phase qubit. Generally, $\sigma$ does not exceed $10\%$ (see Fig.~\ref{fig:SQUID_parameters_draw}(d)) so the two lowest eigenstates of the dc-SQUID Hamiltonian are very close to the two lowest eigenstates of the harmonic oscillator. Consequently, the momentum operator can be expressed in terms of the eigenbasis of the phase qubit as $\widehat{P}_\parallel\approx(|0\rangle\langle1|+|1\rangle\langle0|)/\sqrt{2}=\hat{\sigma}_x^S/\sqrt{2}$ . For $n_g=1/2$, the charge coupling Hamiltonian takes the following form:
\begin{equation}
\widehat{H}_{Coupl,Capa}=-E_{c,c}\,\hat{\sigma}_x^S\hat{\sigma}_x^T.
\end{equation}

The capacitive coupling produces a transverse interaction between the two quantum systems. This coupling was already discussed in Ref.~\cite{Buisson_00,Buisson2003} where a Cooper pair box was coupled to a harmonic oscillator. The capacitive coupling is vanishing in the limit of small $C_2^T$. We notice also that $E_{c,c}$ depends on the bias variables through $\sqrt{\nu_p}$ and therefore is weakly tunable.

\ \\





\subsection{Josephson coupling}
The Josephson coupling Hamiltonian couples the phases $\widehat {X}_\parallel$ and $\hat\psi$ associated with the dc-SQUID and the transistor, respectively. This coupling is mediated via the Josephson junction 2 of the ACPT and reads
\begin{equation}
 \hat H_{Coupl,Jos}=- {E_{J2}^T} \alpha_\parallel
\sqrt{\frac{2E_C^\parallel}{h\nu_p}}\widehat X_\parallel
\sin(\hat\psi-\delta).
\label{eq:H_CouplJosp_general}
\end{equation}
We derive now the expression of the Josephson coupling Hamiltonian for the two different limits of the dc-SQUID. When the dc-SQUID is in the harmonic oscillator limit, the position operator can be written as $\widehat{X}_\parallel=i(\hat a-\hat a^\dag)/\sqrt{2}$. On the other hand, close to $n_g=1/2$, the operator $\sin(\widehat\psi-\delta)$ takes the following form in the charge basis:
$
\sin(\widehat\psi-\delta)=\frac{i}{2}\left(e^{i\delta}|1_{2e}\rangle\langle
0_{2e}|-e^{-i\delta}|0_{2e}\rangle\langle
1_{2e}|\right).\label{eq:sin_psi_charge}$
Using Eq.~(\ref{eq:charge_states}), the Josephson coupling Hamiltonian becomes
\begin{eqnarray}
\widehat{H}_{Coupl,Jos}=&&-iE_{c,j}(\hat a-\hat a^\dag)\big[\sin(\delta/2-\chi)(\sin(\theta)\,\hat\sigma_z^T\nonumber\\
&&+\cos(\theta)\,\hat\sigma_x^T)-\cos(\delta/2-\chi)\,\hat\sigma_y^T\big],
\end{eqnarray}

 where $E_{c,j}=\alpha_\parallel/2E_{J2}^T\sqrt{E_C^\parallel/h\nu_p}$ quantifies the strength of the Josephson coupling. For $n_g=0.5$, we have $\theta=\pi/2$ and the Josephson coupling Hamiltonian reduces to
\begin{eqnarray}
\widehat{H}_{Coupl,Jos}=&&iE_{c,j}(\hat a-\hat a^\dag)(\cos(\delta/2-\chi)\hat\sigma_y^T\nonumber\\
&&-\sin(\delta/2-\chi)\hat\sigma_z^T).
\label{eq:Josephson_coupling_1}
\end{eqnarray}

In the limit of finite anharmonicity $\sigma$, the dc-SQUID can be approximated by a two-level system. In that case, the position operator can be written in the eigenbasis of the phase qubit as $\widehat{X}_\parallel\approx i(|0\rangle\langle1|-|1\rangle\langle0|)/\sqrt{2}=\hat{\sigma}_y^S/\sqrt{2}$. Using the latter expression, the Josephson coupling between the charge and phase qubits takes the following form for $n_g=1/2$:
\begin{eqnarray}
\widehat{H}_{Coupl,Jos}=&&E_{c,j}\cos(\delta/2-\chi)\hat\sigma_y^S\hat\sigma_y^T\nonumber\\&&-E_{c,j}\sin(\delta/2-\chi)\hat\sigma_y^S\hat\sigma_z^T.
\label{eq:Josephson_coupling_2}
\end{eqnarray}

The Josephson coupling contains two different terms. One describes a transverse coupling
$\hat\sigma_y^S\hat\sigma_y^T$ or $(\hat a-\hat a^\dag)\sigma_y^T$ which gives rise to coherent
energy exchange at the resonance between the charge qubit and the phase qubit (or the
oscillator). The effects of this first term on the quantum dynamics of the circuit are
similar to those produced by the capacitive coupling term in
$\hat\sigma_x^S\hat\sigma_x^T$ or $(\hat a-\hat a^\dag)\sigma_x^T$. The second term
$\hat\sigma_y^S\hat\sigma_z^T$ or $(\hat a-\hat a^\dag)\sigma_z^T$  contains a transverse contribution
for the SQUID and a longitudinal term  which depends on the transistor qubit state. Its contribution will explain the quantum measurement of the charge qubit in the
$\nu_S\ll\nu_T$ limit (see Sec.~\ref{sec:read-out}). We notice finally that the two terms
are strongly tunable with the bias parameter $\delta$.



\

\section{Quantum dynamics of the coupled circuit}
The full Hamiltonian of the coupled circuit is given by the sum of the Hamiltonians of the dc-SQUID, the ACPT and the coupling. It reads:
$
\widehat{H}=\widehat{H}_S+\widehat{H}_{ACPT}+\widehat{H}_{Coupl,Capa}+\widehat{H}_{Coupl,Jos}.
$
In order to simplify this Hamiltonian, we consider hereafter the situation where the gate-charge is fixed to $n_g=0.5$. This charge value has been mainly used for the charge qubits experiments (see Refs.~\cite{Nakamura_Nature99,Vion_Science02,Fay2008}). Indeed, at $n_g=0.5$, the charge qubit is insensitive in first order to charge noise and, therefore, its coherence time is longer. We will first derive the expression of the full Hamiltonian when the two quantum systems are close to the resonance condition. We will consider the dc-SQUID either as a phase qubit or as a harmonic oscillator. Finally, we will discuss the quantum measurement scenario of the transistor states. Its principle derives from the coupling to the SQUID, enabling to use it as a detector of the charge qubit's state.

\subsection{A charge qubit coupled to a phase qubit}
We first consider the dc-SQUID as a phase qubit such that the full Hamiltonian governs the quantum dynamics of two-coupled qubits. Using the expressions of the Josephson and capacitive coupling in that limit, we find
\begin{eqnarray}
\widehat{H}=\displaystyle{\frac{h\nu_S}{2}}\hat\sigma_z^S+\displaystyle{\frac{h\nu_T}{2}}\hat\sigma_{z}^T
           -E_{c,c}\,\hat\sigma_x^S\hat\sigma_x^T\nonumber\\+E_{c,j}\cos(\delta/2-\chi)\hat\sigma_y^S\hat\sigma_y^T-E_{c,j}\sin(\delta/2-\chi)\hat\sigma_y^S\hat\sigma_z^T.
\end{eqnarray}
Let us introduce the raising operators $\hat\sigma_+^T$ and $\hat\sigma_+^S$ and the lowering operators $\hat\sigma_-^T$ and $\hat\sigma_-^S$. These operators are defined by
$\hat\sigma_\pm^T\equiv(\hat\sigma_x^T\pm i\hat{\sigma}_y^T)/2$ and
$\hat\sigma_\pm^S\equiv(\hat\sigma_x^S\pm i\hat{\sigma}_y^S)/2$. The terms $\hat\sigma_x^S\hat\sigma_x^T$ and $\hat\sigma_y^S\hat\sigma_y^T$
can be written using the four products
$\hat\sigma_+^{S}\hat\sigma_-^{T}$,
$\hat\sigma_-^{S}\hat\sigma_+^{T}$,$\hat\sigma_+^{S}\hat\sigma_+^{T}$ and
$\hat\sigma_-^{S}\hat\sigma_-^{T}$, whereas the term $\hat\sigma_y^S\hat\sigma_z^T$ is a function of the operators
$\hat\sigma_+^{S}\hat\sigma_z^{T}$ and $\hat\sigma_-^{S}\hat\sigma_z^{T}$.

The product $\hat\sigma_+^{S}\hat\sigma_-^T$ corresponds to an excitation of the phase qubit and de-excitation of the charge qubit. This coupling term mediates the transition between the states $|0,+\rangle$ and $|1,-\rangle$. This transition is only relevant close to resonance, where it contributes to the low-frequency dynamics of the coupled system. Far away from resonance, this term gives rise to high-frequency dynamics (frequencies of the order of $\nu_S$ and $\nu_T$) which is averaged away on the typical time scales of the experiment \cite{Fay2008}. The rotating-wave approximation consists in neglecting these non-resonant terms, called also inelastic terms. The previous reasoning applied to the term $\hat\sigma_-^{S}\hat\sigma_+^T$ leads to the same result. The products $\hat\sigma_+^{S}\hat\sigma_+^T$ and $\hat\sigma_-^{S}\hat\sigma_-^T$
couple the states $|0,-\rangle$ and $|1,+\rangle$. The transition between these two states leads to high-frequency dynamics and are neglected hereafter. For the same reason, the coupling term $\hat\sigma_y^{S}\hat\sigma_z^T$ will be ignored. Finally, close to the resonance, the full Hamiltonian simplifies to
\begin{equation}
\widehat{H}=\displaystyle{\frac{h\nu_S}{2}\,\hat\sigma_z^S+\frac{h\nu_T}{2}\,\hat\sigma_{z}^T-\frac{g}{2}\left(\hat\sigma_+^S\hat\sigma_-^T+\hat\sigma_-^S\hat\sigma_+^T\right)},
\label{eq:total_Hamiltonian_2levels}
\end{equation}
where the coupling strength $g$ is given by $g=2E_{c,c}-2E_{c,j}\cos(\delta/2-\chi)$. We assume hereafter that the gate capacitance $C_g\ll C_2^T$ and $C_1^T\ll C_0$ which is true for the measured circuits in Refs.~\cite{Vion_Science02,Fay2008}. In these limits, we have $C_{n\parallel}\approx2C_0(C_2^T+C_1^T)/(\alpha_\parallel C_2^T)$. The coupling energy can then be written as a function of the capacitance ($\lambda$) and Josephson ($\mu$) asymmetries of the ACPT as:
\begin{equation}
g=\frac{\alpha_\parallel}{2}\sqrt{\frac{E_C^\parallel}{h\nu_p}}\,\left[(1+\lambda)h\nu_p-(1+\mu)E_J^T\cos(\delta/2-\chi)\right].
\label{eq:couplage_g_dvp_before}
\end{equation}
Further simplification of the coupling energy can be obtained using the fact that the charge qubit is described with two charge states, and that the SQUID has a zero anharmonicity. Then, the transistor frequency is related to $\delta$ by $\nu_T=(E_J^T/h)(\cos^2(\delta/2)+\mu^2\sin^2(\delta/2))^{1/2}$ (see Sec.~\ref{sec:ACPT_two_levels}) and we have $\nu_p=\nu_S$. By using these relations, the coupling $g$~(\ref{eq:couplage_g_dvp_before}) can be rewritten as
\begin{eqnarray}
g=\frac{\alpha_\parallel}{2}\sqrt{\frac{E_C^\parallel}{\nu_S}}\,\bigg[(1+& &\lambda)\nu_S-(1+\mu)\nu_T\big(\cos^2(\chi)+\frac{\sin^2(\chi)}{\mu}\big)\bigg].\nonumber\\
\label{eq:couplage_g_dvp}
\end{eqnarray}
Note that coupling $g$ depends strongly on the bias parameters, via the phase $\delta$ and the frequency $\nu_S$. Therefore, the proposed circuit presents an intrinsic tunable coupling between a charge and a phase qubit, which we will analyze in more detail in Sec.~\ref{sec:tunable_coupling}. This transverse coupling enables to realize two-qubit gates operation as for example the $\sqrt{\textrm{iSWAP}}$ operation~\cite{Steffen2006}.
\subsection{A charge-qubit coupled to a tunable harmonic oscillator}
The Hamiltonian (\ref{eq:total_Hamiltonian_2levels}), which describes the dynamics of two coupled qubits, is valid if the second and highest levels of the dc-SQUID do not participate in the quantum dynamics of the circuit. This is the case when the anharmonicity of the SQUID is sufficiently strong. In the limit of zero anharmonicity, the quantum dynamics of the coupled circuit is described by the Jaynes-Cummings Hamiltonian~\cite{Jaynes-Cummings}
\begin{equation}
\widehat{H}=\displaystyle{h\nu_p{(\hat a\hat a^\dag+\frac{1}{2})}+\frac{h\nu_T}{2}\,\hat\sigma_{z}^T-\frac{g}{2}\left(\hat a^\dag\hat\sigma_-^T+\hat a\hat\sigma_+^T\right)}.
\label{eq:total_Hamiltonian_multilevels}
\end{equation}

This Hamiltonian (\ref{eq:total_Hamiltonian_multilevels}) is very similar to the one obtained by coupling a charge~\cite{Blais2004} or a transmon~\cite{Koch_PRA07} or phase qubit~\cite{Sillanpaa_Nature07} to a coplanar waveguide cavity. In our circuit the resonator is realized by a micro dc-SQUID. It can be viewed as "micro-resonator" more convenient for integration than usual coplanar resonators since its size ranges in the micrometer scale three orders of magnitude smaller. Moreover its resonance frequency is strongly tunable. However up to now it suffers from a much shorter coherence time.




%

\subsection{Quantum measurements of the charge qubit by the dc-SQUID}
\label{sec:read-out}

Our theoretical analysis introduces two different kinds of coupling which can be used to perform quantum measurements of the charge qubit by the dc-SQUID. Here we will apply this analysis to the quantronium read-out \cite{Vion_Science02} and the adiabatic quantum transfer method used in Ref.~\cite{Fay2008}.

\subsubsection{Quantronium read-out}

\begin{figure}
\centering
\includegraphics[width=0.73\linewidth]{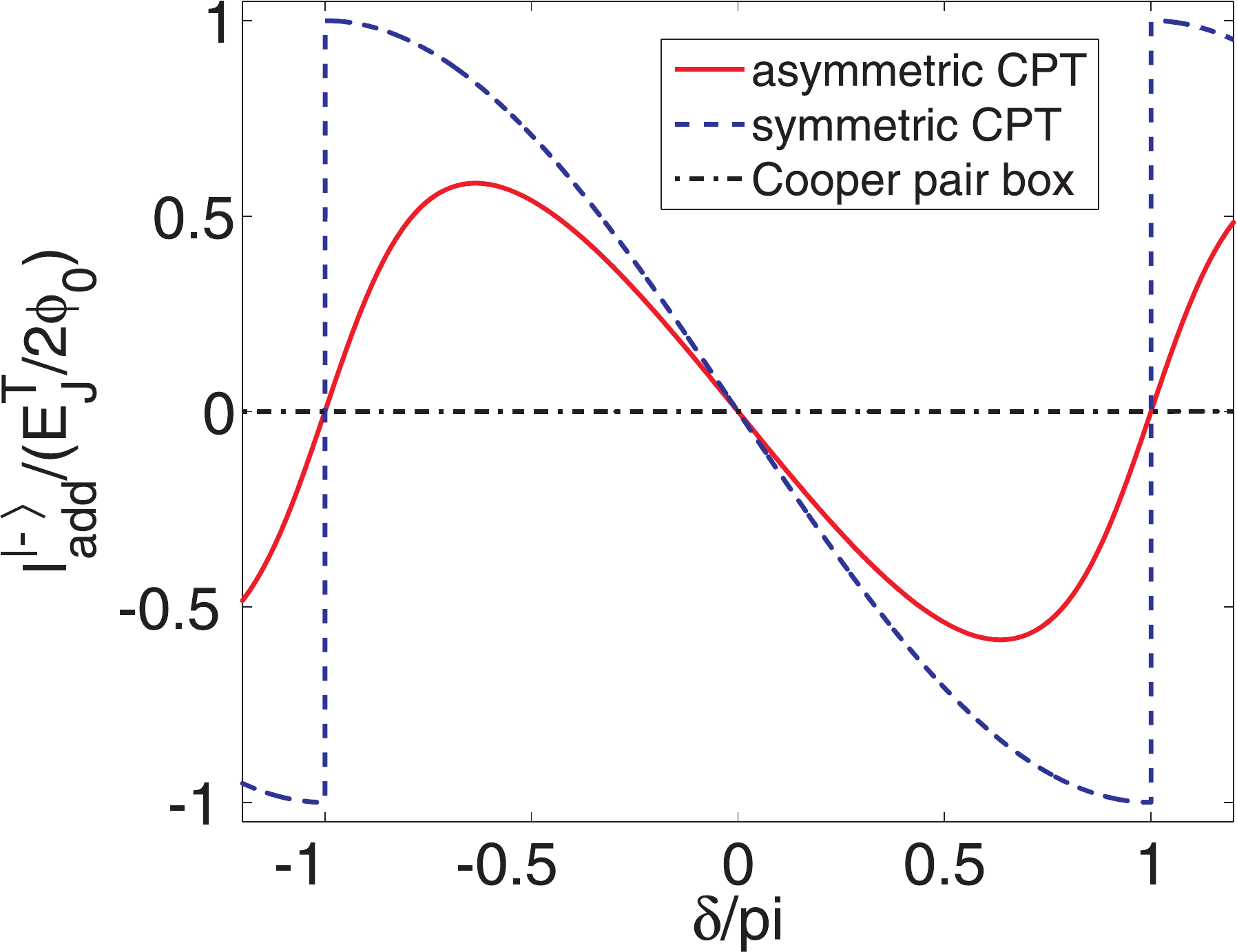}
\caption{(Color online). Effective current $I_{add}^{|-\rangle}$ at $n_g=1/2$ as a function of the phase $\delta$ for an asymmetric ($\mu=0.42$) and symmetric ($\mu=0$) transistors and a Cooper pair box ($\mu=1$).}
\label{fig:IS_vs_delta}
\end{figure}

This read-out is obtained the limit where $\nu_T\gg\nu_S$.
As we will see, even if the qubits are far from resonance the coupling still affects the dynamics of the circuit. Indeed, since the dynamics of the ACPT is much faster than the dc-SQUID dynamics, the dc-SQUID is only sensitive to the average value of the ACPT operators. After some straightforward algebra, the effective Hamiltonian of the dc-SQUID takes the following form:
\begin{equation}
\hat H_{S,eff}^{|\pm\rangle}=\widehat H_S-\frac{C_T}{C_{n\parallel}} \frac{\partial E_{|\pm\rangle}}{\partial n_g}\widetilde{P}_{\parallel}+\alpha_\parallel\frac{\partial E_{|\pm\rangle}}{\partial \delta}\widetilde{X}_{\parallel},
\end{equation}
with $E_{|+\rangle}$ and $E_{|-\rangle}$ the eigenenergies of the ACPT associated with the states $|+\rangle$ and $|-\rangle$. For $n_g=0$ or $n_g=1/2$, we have ${\partial E_{|\pm\rangle}}/{\partial n_g}=0$ and the SQUID Hamiltonian simplifies to
\begin{equation}
\hat H_{S,eff}^{|\pm\rangle}=\widehat H_S-\phi_0 I_{add}^{|\pm\rangle}\widetilde X_{\parallel},
\end{equation}
where $I_{add}^{|\pm\rangle}\equiv -(\alpha_\parallel/\phi_0) \partial E_{|\pm\rangle}/\partial \delta$  adds to the bias current $I_b$ and depends on the state of the ACPT $|-\rangle$ or $|+\rangle$. For $n_g=1/2$, in the limit of two charge states only, the additional current reads
\begin{equation}
I_{add}^{|\pm\rangle}(n_g=1/2)=\pm\frac{E_J^T}{2\phi_0}(1+\mu)\sin({\delta}/{2} -\chi).
\label{eq:currents_add}
\end{equation}
The currents $I_{add}^{|+\rangle}(n_g=1/2)$ and $I_{add}^{|-\rangle}(n_g=1/2)$ take opposite values and, therefore, can be used to determine the transistor state. Fig.~\ref{fig:IS_vs_delta} shows the normalized current $I_{add}^{|-\rangle}/(E_J^T/2\phi_0)$ for $n_g=1/2$ as a function of the phase $\delta$ for an asymmetric ($\mu=0.42$) and a symmetric ($\mu=0$) transistor and a Cooper pair box ($\mu=1$). For a symmetric transistor, the current has a strong discontinuity at $\delta=\pm\pi$ jumping between the two extreme values $-E_J^T/2\phi_0$ and $E_J^T/2\phi_0$. This discontinuity disappears for a finite Josephson asymmetry; its maximum value is lower. Naturally, for a Cooper pair box, the additional current is zero as one of the junctions is replaced by a pure capacitance.

\begin{figure}
\centering
\includegraphics[width=0.73\linewidth]{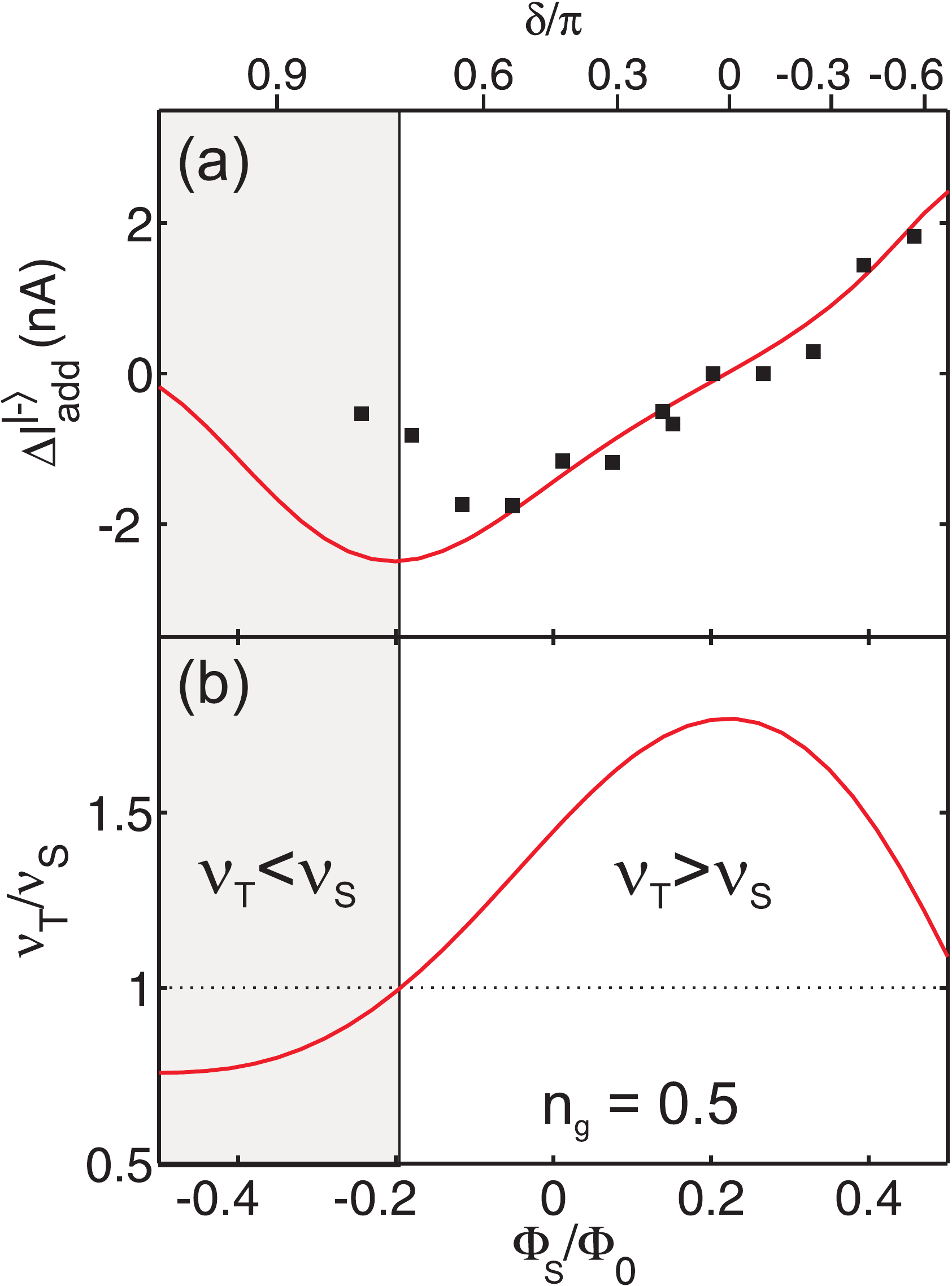}
\caption{(Color online). (a) Experimental (black squares) and theoretical (red line) current difference $\Delta I_{add}^{|-\rangle}\equiv I_{add}^{|-\rangle}(n_g=1/2)-I_{add}^{|-\rangle}(n_g=0)$ as a function of the flux $\Phi_S$ (or the phase $\delta$ indicated in the upper scale). (b) Frequencies ratio $\nu_T/\nu_S$ as a function of $\Phi_S$.}
\label{fig:addcurrent_vsdelta}
\end{figure}

As the switching probability of the dc-SQUID depends strongly on the bias current \cite{Claudon2007} the additional current $I_{add}$ can be used to detect the ACPT state. For instance, in the Quantronium circuit \cite{Vion_Science02}, the state of a symmetric Cooper Pair Transistor (CPT) is read out by measuring the switching probability of a Josephson junction placed in parallel with the CPT and which replaces the dc-SQUID in the studied circuit. In that case, the charge qubit read-out is explained by the $\widehat X_\parallel^S\hat\sigma_z^T$ term resulting from the Josephson coupling (\ref{eq:H_CouplJosp_general}).

The expression (\ref{eq:currents_add}) of the additional current has been established under the condition $\nu_T\gg\nu_S$. Is this expression still valid when $\nu_T\approx\nu_S$? In order to answer this question, we compare experimental data and theoretical predictions for the difference of the additional currents at $n_g=1/2$ and $n_g=0$ in the ground state of the ACPT. This difference noted $\Delta I_{add}^{-}\equiv I_{add}^{|-\rangle}(n_g=1/2)-I_{add}^{|-\rangle}(n_g=0)$ is shown in Fig.~\ref{fig:addcurrent_vsdelta} as a function of the flux $\Phi_S$.
The theoretical curve calculated without any free parameters agrees well with the experiment when $\nu_T/\nu_S\gg1$, but differs from the experiment when $\nu_T\approx\nu_S$. The measured current amplitude drops when $\nu_T$ is close to $\nu_S$ suggesting a drop in the contrast of the quantum measurement when $\nu_T\approx\nu_S$.

\subsubsection{Adiabatic quantum transfer}

\begin{figure}
\centering
\includegraphics[width=0.90\linewidth]{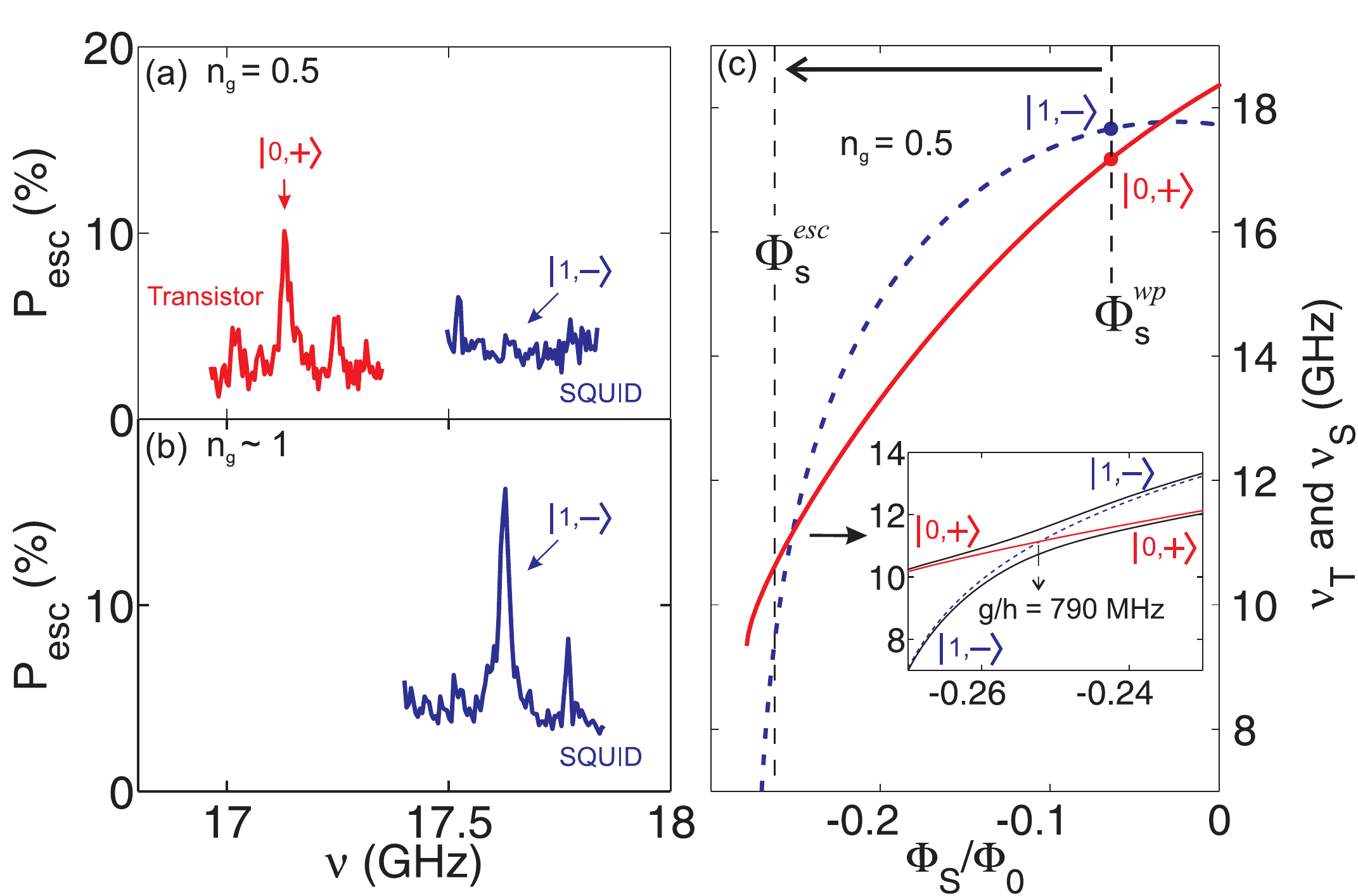}
\caption{(Color online). Spectroscopies of the phase (blue curve) and charge (red curve) qubit studied in Ref. \cite{Fay2008} at (a) $n_g=0.5$ and (b) $n_g\approx 1$ ($\nu_T\gg\nu_S$) for the working point ($I_b^{wp}=1957$ nA,$\Phi_S^{wp}=-0.064\ \Phi_0$). These spectroscopies are measured with a nanosecond flux pulse which changes the flux in the SQUID loop from $\Phi_S^{wp}$ to $\Phi_S^{esc}$.  (c) Evolution of the qubit frequencies $\nu_T$ and $\nu_S$ during this pulse. Blue and red points indicate the frequencies $\nu_S$ and $\nu_T$ at the working point of spectroscopy (a). Inset shows an antilevel crossing in the energy spectrum where an adiabatic transfer happens during the flux pulse, the state $|1,-\rangle$ ($|0,+\rangle$) being transferred to the state $|0,+\rangle$ ($|1,-\rangle$).}
\label{fig:Adiabatic_transfert}
\end{figure}

In Fig.~\ref{fig:Adiabatic_transfert}, the first energy levels of the charge and phase qubit in the coupled circuits are plotted. The quantum measurement of the charge qubit is performed by a nanosecond flux pulse which transfers the quantum state $|0,+\rangle$ prepared at the working point (I$_b^{wp}$,$\Phi_S^{wp}$) to the measurement point (I$_b^{esc}$,$\Phi_S^{esc}$). At that point the SQUID is very close to the critical line and the escape probability is finite. Spectroscopy measurements show clearly the read-out of the charge qubit by this method even in the limit $\nu_T\approx\nu_S$. How can we explain this read-out?

The coupling terms $\hat\sigma_+^{S}\hat\sigma_-^T$ and $\hat\sigma_-^{S}\hat\sigma_+^T$ produce an anti-level-crossing the amplitude of which depends on the coupling strength $g$ (see Inset of Fig.~\ref{fig:Adiabatic_transfert}(c)). During a nanosecond flux pulse, the coupled system remains in its original energy state; as a result, the quantum state $|0,+\rangle$ evolves adiabatically into to the state $|1,-\rangle$. Due to large value  of the coupling strength close to the escape point, Landau-Zener transitions can be neglected. The final state, i.e. $|1,-\rangle$ or $|0,-\rangle$, is determined by the switching measurement.

\section{Tunable Coupling}
\label{sec:tunable_coupling}

In this section, we calculate numerical values of the coupling strength $g$, using the circuit parameters of Ref.~\cite{Fay2008} and show that the coupling can be tuned over a wide range with the bias parameters.

\begin{figure}
\centering
\includegraphics[width=0.8\linewidth]{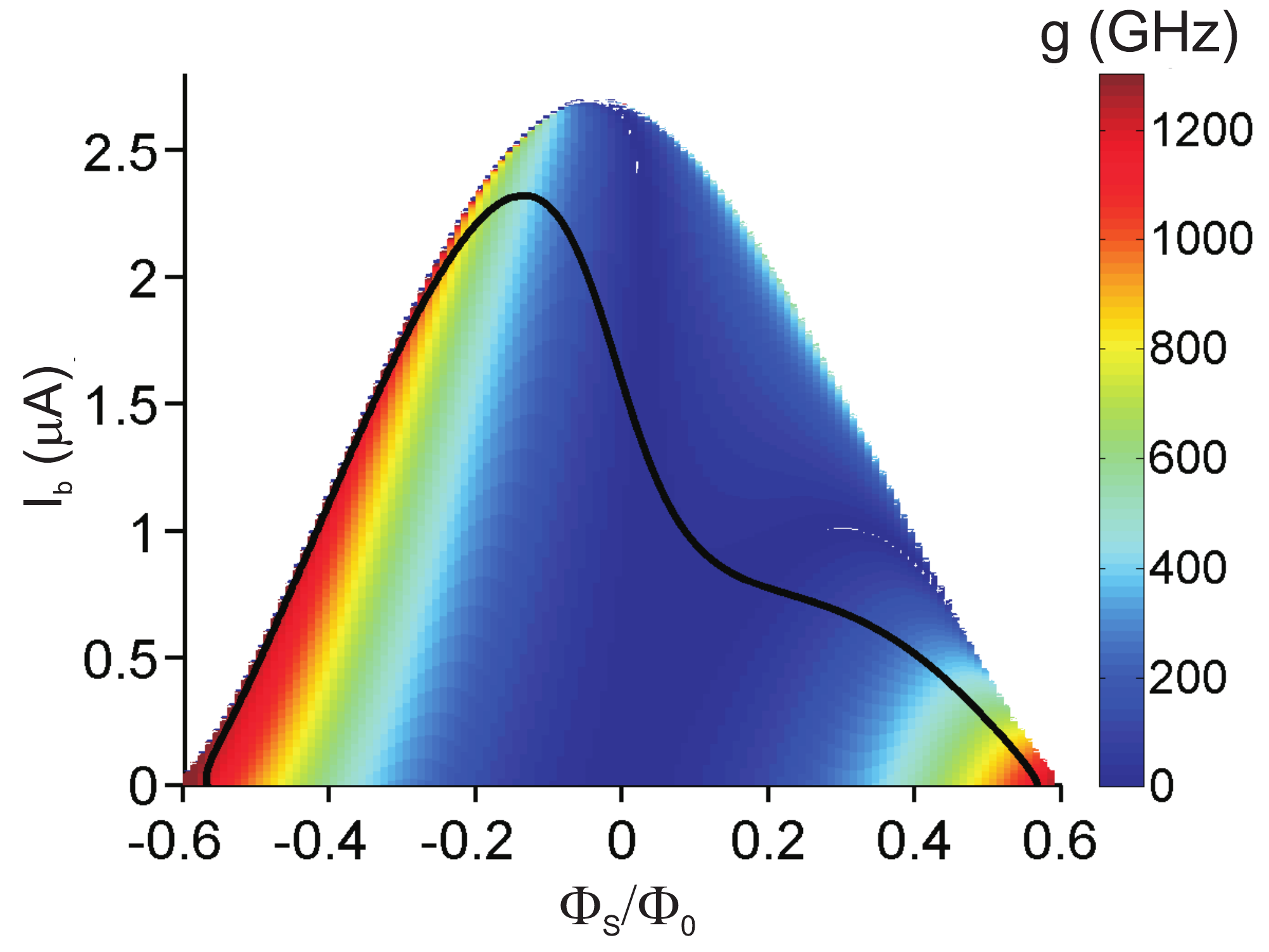}
\caption{(Color online). Color plot of the analytical coupling~(\ref{eq:couplage_g_dvp}) as a function of the bias current $I_b$ and the flux $\Phi_S$, for the wells family [0] of the dc-SQUID. The solid line indicates where the two qubits are in resonance ($\nu_S=\nu_T$).}
\label{fig:coupling_arche0}
\end{figure}

Fig.~\ref{fig:coupling_arche0} presents the dependence of the coupling $g$ as a function of the bias current $I_b$ and the flux $\Phi_S$ for the family [0] of the dc-SQUID. We find that the coupling between the two qubits can be tuned from zero to more than 1.2 GHz. This on-off coupling is one of the needed requirements to realize ideally one- and two-qubit gates. In Ref.~\cite{Fay2008}, the coupling $g$ has been measured at resonance, where the coupling effect on the qubits is maximal. Far away from resonance, the eigenenergies of each individual qubit are shifted by the amount $g^2/4|\Delta|$ where $\Delta=h\nu_S-h\nu_T$ is the detuning. So, if the detuning is large the coupling can be neglected. The solid line of Fig.~\ref{fig:coupling_arche0} shows where the qubits are in resonance and, consequently, where the coupling can be easily measured by spectroscopy.

\begin{figure}
\centering
\includegraphics[width=0.8\linewidth]{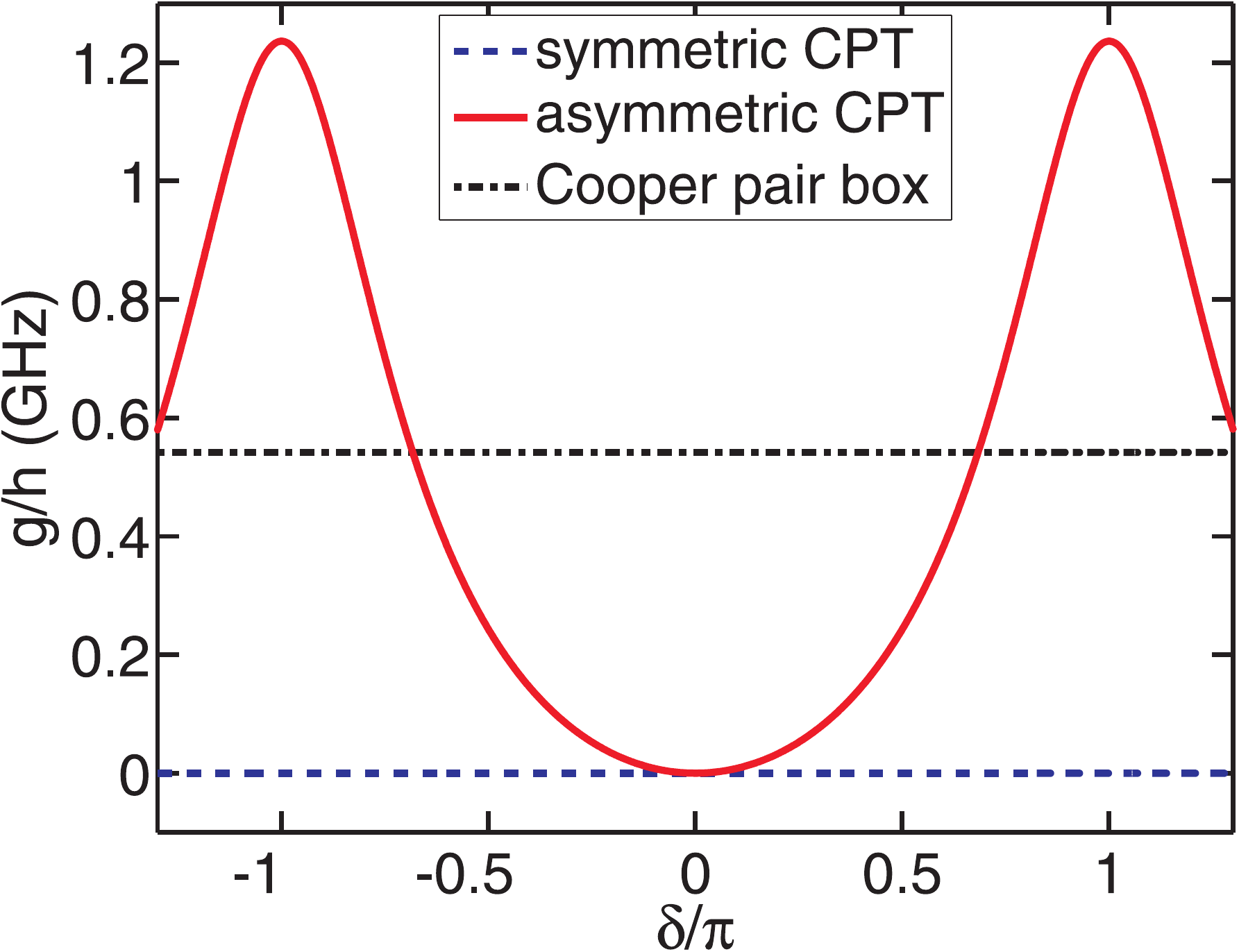}
\caption{Coupling $g$ derived from Eq. (\ref{eq:couplage_g_dvp}) as a function of $\delta$ when the two qubits are in resonance ($\nu_T=\nu_S$) for three different charge qubits: (red curve) an asymmetric transistor with $\lambda=\mu=-41.6\%$, (bue curve) a symmetric transistor $\lambda=\mu=0$ and (black curve) a Cooper pair box with $\lambda=-41.6\%$ and $\mu=-1$. The calculations have been realized by using the parameters of Ref. \cite{Fay2008} given in Appendix~\ref{appendix:parameters}.}
\label{fig:coupling_vs_delta_diffchargequbits}
\end{figure}

Fig. \ref{fig:coupling_vs_delta_diffchargequbits} shows the dependence of the coupling $g$ as a function of the phase $\delta$ for three different charge qubits: an asymmetric transistor \cite{Fay2008}, a symmetric transistor \cite{Vion_Science02} and a Cooper pair box \cite{Nakamura_Nature99}. The coupling has been calculated using the expression (\ref{eq:couplage_g_dvp}) at the resonance $\nu_S=\nu_T$ with the parameters of Ref.~\cite{Fay2008} and only the ACPT asymmetries have been varied. For the asymmetric transistor with $\lambda=\mu=-41.6\%$, the coupling $g$ (red curve) is maximum at $\delta=\pm\pi$ where it equals to $1217$~MHz; it becomes zero at $\delta=0$. In the case of the {\em symmetric} transistor with $\lambda=\mu=0$, the coupling strength reads $g={\alpha_\parallel}/{2}\sqrt{{E_C^\parallel}/{\nu_S}}\left(\nu_S-\nu_T\right)$. Consequently, at the resonance, the coupling between a {\em symmetric} transistor and a SQUID is zero. For a Cooper pair box with $E_{J2}^T=0$ ($\mu=-1$ and $\lambda=-41.6\%$), the coupling reads
$g=\frac{\alpha_\parallel}{2}\sqrt{{E_C^\parallel}{h\nu_S}}(1+\lambda)$. This corresponds to the result of Ref.~\cite{Buisson2003}. The calculated coupling $g$ (black curve) does not depend on $\delta$ and remains equal to $514$~MHz.

\begin{figure}
\centering
\includegraphics[width=0.9\linewidth]{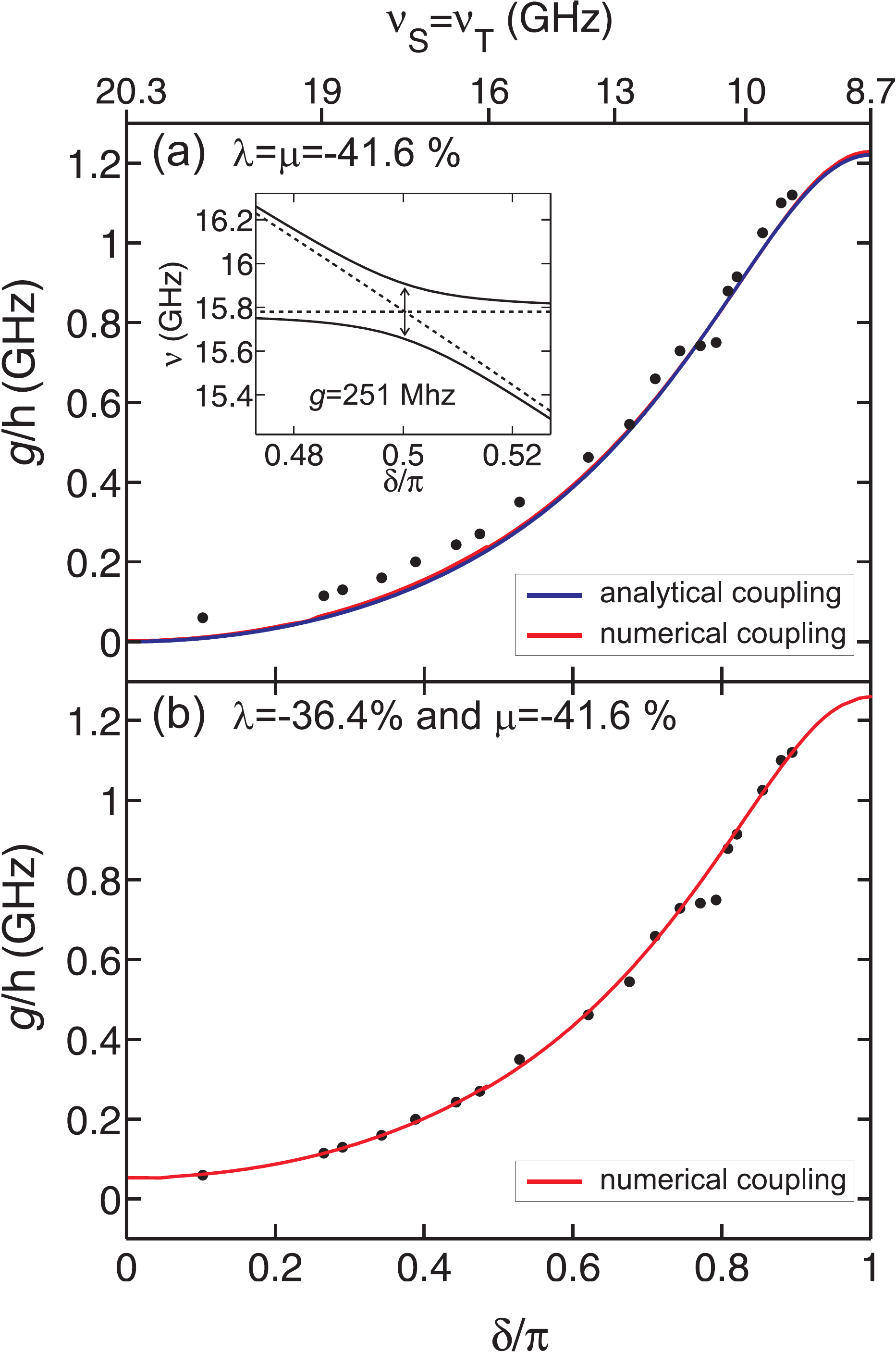}
\caption{(Color online). (a) Coupling $g$ as a function of $\delta$ when the SQUID and the ACPT are in resonance ($\nu_T=\nu_S$). The black points are the experimental couplings measured in Ref.~\cite{Fay2008}. The blue line is the theoretical coupling calculated from the analytical expression (\ref{eq:couplage_g_dvp}), using the circuit parameters of Ref.~\cite{Fay2008} and equal Josephson and capacitance asymmetries $\mu=\lambda=-46.6$ $\%$. The red line is the numerical coupling for the same circuit parameters. It is calculated by diagonalizing the full Hamiltonian in the basis of 8 charge states and the first 9 harmonic dc-SQUID states. The inset shows the numerical simulations of the eigenenergies of the circuit with (full curve) and without (dashed curve) coupling from which we extract the coupling $g/h=251$~MHz at $\delta=0.5\pi$. (b) Numerical coupling $g$ as a function of $\delta$ when the SQUID and the ACPT are in resonance, calculated using the circuit parameters used in (a)  but with a different capacitance asymmetry $\lambda=0.87\mu=-36.4\%$.}
\label{fig:coupling_vs_delta_exp}
\end{figure}

Fig.~\ref{fig:coupling_vs_delta_exp}(a) shows the dependence of the analytical and numerical couplings at the resonance as a function of the absolute value of the phase $\delta$. The numerical simulations allow us to check the validity of the analytical coupling. These simulations have been realized by diagonalizing the full Hamiltonian in the basis of 20 charge states and the first 9 excited states of the dc-SQUID in the absence of anharmonicity. The dc-SQUID and ACPT Hamiltonians, as well as the Josephson and capacitive coupling Hamiltonians can all be expressed naturally in these bases. The numerical coupling is found by calculating the energy spectrum as a function of $\delta$ between $-\pi$ and $\pi$ for a fixed frequency $\nu_S$. The spectrum is first calculated without the coupling terms in order to find the two opposite values of $\delta=\pm\delta_r$ where the ACPT and the dc-SQUID are in resonance ($\nu_S=\nu_T$), i.e. where the second and third energy bands intersect. In the presence of coupling, the degeneracy between the eigenstates is lifted and an anti-level crossing appears with an energy separation equal to the coupling $g$ at $\delta=\pm\delta_r$. As an example, the inset of Fig.~\ref{fig:coupling_vs_delta_exp}(a) shows the energy spectrum for $\delta_r=0.5\pi$ with and without coupling from which we extract $g/h=251$~MHz. The numerical and analytical simulations remarkably give quite close results which confirms the validity of the analytical expression (\ref{eq:couplage_g_dvp}). The theoretical coupling is calculated here without any free parameters by assuming equal Josephson and capacitive asymmetries ($\lambda=\mu=-41.6\%$). Finally, we note the good agreement with the experimental coupling measured in Ref.~\cite{Fay2008} and shown in black points in Fig.~\ref{fig:coupling_vs_delta_exp}. By adjusting the capacitive asymmetry to $\lambda=-36.4\%$ the numerical coupling is in perfect agreement with the experiment as shown in Fig.~\ref{fig:coupling_vs_delta_exp}(b).

\section{Conclusion}
In conclusion, we have analyzed in detail the theoretical quantum dynamics of a superconducting circuit based on a dc-SQUID in parallel to an Asymmetric Cooper Pair Transistor (ACPT). The Lagrangian of the circuit was first established from the current conservation equations expressed at each node of the circuit. The Hamiltonian, deduced from the Lagrangian, is decomposed in three distinct terms, namely the dc-SQUID, the ACPT and the coupling Hamiltonians. We first studied the individual dynamics of the dc-SQUID and the ACPT. Depending on its anharmonicity, the dc-SQUID can be seen either as a harmonic oscillator or as a phase qubit, whereas the ACPT behaves as a charge qubit. In addition to the optimal bias points ($\delta=0,n_g=1/2$) which was successfully demonstrated in a symmetric Cooper pair transistor, the ACPT presents a second optimal point ($\delta=\pi,n_g=1/2$). At these points, the charge qubit is insensitive in first order to the charge, flux and current noise and therefore shows a larger coherence time. We found that the coupling Hamiltonian between the dc-SQUID and the ACPT is made of two different terms corresponding to the Josephson and the capacitive couplings, which mix phases and charges of both sub-circuits, respectively.

The Hamiltonian of the full circuit was discussed through two different limits of the dc-SQUID. When it behaves as a harmonic oscillator, the quantum dynamics of the circuit is described by the well-known Jaynes-Cummings Hamiltonian. The microscopic circuit is then similar to a two-level atom coupled to a single-mode optical cavity. It offers compared to this latter a better tunability, a faster control and read-out of the quantum system, and a good scalability for complex achitectures implementation. For example a circuit of several ACPT qubit in parallel could be considered whose quantum information will be mediated by the dc-SQUID~\cite{Plastina_PRB03}. When the anharmonicity of the dc-SQUID is strong, it behaves as a phase qubit. The full circuit is then described as the coupling of two different class of qubits, i.e., a phase and a charge qubit. The coupling Hamiltonian contains terms in $\hat\sigma_y^T\hat\sigma_y^S$ and $\hat\sigma_x^T\hat\sigma_x^S$ which are prominent when the two qubits are in resonance. These terms allow two-qubits gate operations as the $\sqrt{\textrm{iSWAP}}$ gate. In addition they enable to read out the quantum state of the ACPT by a nanosecond flux pulse as observed in Ref.\cite{Fay2008}. Indeed such a pulse produces an adiabatic quantum transfer of the state $|0,+\rangle$ into the state $|1,-\rangle$, i.e. the energy quantum stored in the ACPT is transferred into the dc-SQUID in order to be detected. A non-resonant term in $(\hat a-\hat a^{\dag})\hat\sigma_z^T$ or in $\hat\sigma_y^S\hat\sigma_z^T$ is present in the Josephson coupling. Although its effect on the quantum dynamics is weak, this latter term explains the charge qubit read-out in the limit $\nu_T\gg\nu_S$ by mean of an effective additional current in the dc-SQUID. That read-out method is employed in the Quantronium circuit \cite{Vion_Science02}.

 In both limits of the dc-SQUID, we demonstrated that the coupling can be strongly tuned, mainly with the Josephson coupling term which has a strong phase $\delta$ dependence. It can be used to accomplish two-qubit gate operations, and can also be turned off in order to perform one-qubit gate operations without desturbing the unaddressed qubit.

\section*{Acknowledgment}

We thank J. Claudon, E. Hoskinson, L. L\'evy, and D.
Est\`eve for fruitful discussions. This work was supported
by the EuroSQIP, MIDAS and SOLID european projects,
by the french ANR'QuantJO' and by Institut Universitaire
de France.

\
\newpage
\
\ \\
\newpage
\
\\
\appendix

\section{Current conservation laws}
\label{appendix:current_conservation} The current conservation law, applied to each node
of the circuit (see Fig. \ref{fig:schematic_circuit}(b)), yields six equations for the
active phases $\varphi_1$, $\varphi_2$, $\psi$, $\theta$ and $\xi$. These equations are
identical to the Euler-Lagrange equations derived from the circuit Lagrangian,
\eqref{eq:Lagrangien_T-V}, \eqref{eq:Lagrangien_T}, \eqref{eq:Lagrangien_V} and read,
respectively,

%
\begin{eqnarray}
&&\phi_0\frac{\xi-\varphi_1}{L_1}-C_0\phi_0\ddot{\varphi_1}-I_0\sin(\varphi_1)=0; \\
&&\phi_0\frac{\gamma-\varphi_2}{L_2}-C_0\phi_0\ddot{\varphi_2}-I_0\sin\varphi_2=0;\\
&&I_2^T\sin(\xi-\psi-\phi_T)+C_2^T\phi_0(\ddot{\xi}-\ddot{\psi})\nonumber\\
&&\qquad-C_g\phi_0(\ddot{\psi}-\ddot{\theta})-C_1^T\phi_0\ddot{\psi}-I_1^T\sin(\psi)=0;\\
&&C_g\ddot{\psi}-(C_P+C_g)\ddot{\theta}=0;\\
&&\phi_0\frac{\gamma-\xi-\phi_S}{L_3}-\phi_0\frac{\xi-\varphi_1}{L_1}\nonumber\\
&&\qquad-C_2^T\phi_0(\ddot\xi-\ddot\psi) -I_2^T \sin(\xi-\psi-\phi_T)=0. \label{eq:phase_xi}
\end{eqnarray}
In Sec.~\ref{subsection:current_conservation}, we show that this system of six equations
can be reduced to four equations by ignoring the high frequency quantum dynamics of the
phases $\gamma$ and $\xi$.

\section{Parameters of the coupled circuit studied in Ref. \cite{Fay2008}}
\label{appendix:parameters} Throughout this article, we illustrate the theory with
numerical values and plots calculated by using the parameters of the circuit studied in
Ref.~\cite{Fay2008}. These parameters are collected in Tab.~\ref{tab:parameters}.
\begin{table}[htb!]
\centering
\vspace{1cm} \centering
\begin{tabular}{l|c|c}
\hline\hline
                                  & Label  &     Value     \\
\hline
{Parameters of the dc-SQUID}           &           &                              \\
%
\hspace{0.4cm} {Critical current}  & \multirow{2}{*}{$I_0$}     & \multirow{2}{*}{$1.356 \: \mu\text{A}$} \\[-0.03in]
\hspace{0.4cm} {per Josephson junction (JJ)}  &    & \\
\hspace{0.4cm} {Capacitance per JJ}           & $C_0$     & $227 \: \text{fF}$        \\
\hspace{0.4cm} {Loop inductance}           & $L_s$     & $190 \: \text{pH}$        \\
\hspace{0.4cm} {Inductance asymmetry}          & $\eta$    & $0.28$                           \\
\hspace{0.4cm} {Bidimensionality parameter}             & $b$        & $1.28$               \\
\hline
{Parameters of the ACPT}           &           &                                     \\
\hspace{0.4cm} {Critical current of the first JJ}   & $I_1^T$     & $30.1 \: \text{nA}$  \\
\hspace{0.4cm} {Critical current of the second JJ}            & $I_2^T$     & $12.4\: \text{nA}$        \\
\hspace{0.4cm} {Capacitance of the first JJ}            & $C_1^T$     & $2.0 \: \text{fF}$         \\
\hspace{0.4cm} {Capacitance of the second JJ}          & $C_2^T$    & $0.9 \: \text{fF}$                          \\
\hspace{0.4cm} {Critical current asymetry} & $\mu$ & $-41.6\,\%$\\
\hspace{0.4cm} {Capacitance asymetry} & $\lambda$ & $-37.7\,\%$\\
\hspace{0.4cm} {Gate capacitance}         & $C_g$    & $29$ aF                          \\

\hline
\hline
\end{tabular}
\caption{Electric parameters of the coupled circuit studied by Fay \textit{et al.} \cite{Fay2008}.}
\label{tab:parameters}
\end{table}

\section{Conjugate variables}
\label{appendix:conjugate_variables}
The phases $(X_\parallel,Y_\perp,\psi,\theta)$ and their conjugate momenta $(-\hbar
P_\parallel,-\hbar P_\perp,-\hbar n,-\hbar n_Q)$ are the appropriate variables of the
circuit Hamiltonian (\ref{eq:Hamiltonien_total}). The momenta are related to the
velocities$(\dot X_\parallel,\dot Y_\perp,\dot \psi,\dot\theta)$ involved in the kinetic
part of the Lagrangian (\ref{eq:lagrangien_cinetique}) by the following expressions
\cite{Landau}:

\begin{eqnarray}
-\hbar P_\parallel &=& \frac{\partial {\mathcal
L}}{\partial\dot{X}_\parallel} =
\phi_0^2\left\{(2C_0+\alpha_\parallel^2 C_2^T)\dot
X_\parallel-\alpha_\parallel C_2^T\left(\alpha_\perp\dot{Y}_\perp+\dot{\psi}\right)\right\}\nonumber\\
\ \\
-\hbar P_\perp &=& \frac{\partial {\mathcal
L}}{\partial\dot{Y}_\perp} = \phi_0^2\left\{(2C_0+\alpha_\perp^2
C_2^T)\dot
Y_\perp-\alpha_\perp C_2^T \left(\alpha_\parallel\dot{X}_\parallel-\dot{\psi}\right)\right\}\nonumber\\
\ \\
-\hbar n &=& \frac{\partial {\mathcal L}}{\partial\dot{\psi}} =
\phi_0^2\left\{
C_\Sigma\dot\psi-C_2^T(\alpha_\parallel\dot{X}_\parallel-\alpha_\perp\dot{Y}_\perp)-C_g\dot{\theta}\right\}\label{eq:charge_ile}\\
-\hbar n_Q &=& \frac{\partial {\mathcal L}}{\partial\dot{\theta}}
=\phi_0^2\left\{(C_P+C_g)\dot{\theta}-C_g\dot{\psi}\right\}.\label{eq:nq}
\end{eqnarray}

\section{Pauli matrixes}
\label{annexe:Pauli_matrix}

The Pauli matrices related to the dc-SQUID are defined in the eigenbasis of the phase
qubit $\{|1\rangle,|0\rangle\}$ as
\begin{equation}
\sigma_z^S=
 \left(
  \begin{array}{cc}
    1 & 0 \\
    0 & -1 \\
  \end{array}
\right) ,\ \
 \sigma_x^S=
 \left(
  \begin{array}{cc}
    0 & 1 \\
    1 & 0 \\
  \end{array}
\right)
 ,\ \
 \sigma_y^S=
 \left(
  \begin{array}{cc}
    0 & -i \\
    i & 0 \\
  \end{array}
\right).
\end{equation}
Similarly the Pauli matrices related to the ACPT ($\sigma_z^T$, $\sigma_x^T$, $\sigma_y^T$) are defined in the eigenbasis of the charge qubit
$\{|+\rangle,|-\rangle\}$.


\end{document}